\documentclass[twocolumn,preprintnumbers,amsmath,amssymb]{revtex4}

\usepackage{graphicx}
\usepackage{dcolumn}
\usepackage{bm}
\usepackage{float}

\usepackage{multibib}
\newcites{Methods}{Methods References}
\newcites{Supplement}{Supplement}

\usepackage{amssymb}
\usepackage{color}
\usepackage{soul}

\newcommand{\ket}[1]{\vert #1 \rangle}

\renewcommand{\figurename}{Figure}

\begin{document}

\title{Wavelength-scale errors in optical localization due to spin–orbit coupling of light}

\author{G. Araneda,$^{1,}$\footnote{These authors contributed equally to this work}$^,$\footnote{Electronic address: \texttt{gabriel.araneda-machuca@uibk.ac.at}}
S. Walser,$^{2,*}$
Y. Colombe,$^1$
D. B. Higginbottom,$^{1,3}$
J. Volz,$^{2,}$\footnote{Electronic address: \texttt{jvolz@ati.ac.at}}
R. Blatt,$^{1,4}$ and
A. Rauschenbeutel$^{2,}$\footnote{Electronic address: \texttt{arno.rauschenbeutel@ati.ac.at}}
}

\affiliation{$^1$Institut f\"{u}r Experimentalphysik, Universit\"{a}t Innsbruck, Technikerstra\ss e 25, 6020 Innsbruck, Austria}
\affiliation{$^2$Vienna Center for Quantum Science and Technology, TU Wien-Atominstitut, Stadionallee 2, 1020 Vienna, Austria}
\affiliation{$^3$Centre for Quantum Computation and Communication Technology, Research School of Physics and Engineering, The Australian National University, Canberra ACT 0200, Australia}
\affiliation{$^4$Institut f\"{u}r Quantenoptik und Quanteninformation, \"{O}sterreichische Akademie der Wissenschaften, Technikerstra\ss e 21a, 6020 Innsbruck, Austria}

\maketitle

\textbf{The precise determination of the position of point-like emitters and scatterers using far-field optical imaging techniques is of utmost importance for a wide range of applications in medicine, biology, astronomy, and physics \cite{boas2011handbook,kovalevsky2004fundamentals,novotny_hecht_2006}. Although the optical wavelength sets a fundamental limit to the image resolution of unknown objects, the position of an individual emitter can in principle be estimated from the image with arbitrary precision. This is used, \emph{e.g.}, in stars' position determination~\cite{anderson2000toward} and in optical super-resolution microscopy~\cite{hell2007far}. Furthermore, precise position determination is an experimental prerequisite for the manipulation and measurement of individual quantum systems, such as atoms, ions, and solid state-based quantum emitters~\cite{Alberti2016,monroeadaptative,sapienza2015nanoscale}. Here we demonstrate that spin-orbit coupling of light in the emission of elliptically polarized emitters can lead to systematic, wavelength-scale errors in the estimate of the emitter's position.
Imaging a single trapped atom as well as a single sub-wavelength-diameter gold nanoparticle, we demonstrate a shift between the emitters' measured and actual positions which is comparable to the optical wavelength. Remarkably, for certain settings, the expected shift can become arbitrarily large. Beyond their relevance for optical imaging techniques, our findings apply to the localization of objects using any type of wave that carries orbital angular momentum relative to the emitter's position with a component orthogonal to the direction of observation.
}

An ideal imaging system with aperture diameter $D$ has an angular resolution $\lambda/D$ where $\lambda$ is the wavelength of the imaging light. Objects with smaller angular diameter cannot be resolved and produce an image given by the point-spread function (PSF) of the optical system. In spite of this so-called diffraction limit, fitting the PSF to the image allows one to estimate its position with a precision that exceeds the diffraction limit, limited only by the image's signal to noise ratio~\cite{thompson2002}. The central assumption of this method is that the emitters' positions in the object plane correspond to the centroid of the PSF measured in the image plane, provided that the optical system is focussed.

It is known that the centroid of the image can be affected by imperfect focussing when the emission pattern of the object is anisotropic, as for a linear dipole. Depending on the orientation of the latter, this may lead to lateral shifts of a few tens of nanometres, \emph{i.e.}, much smaller than the diffraction limit~\cite{Enderlein2006,Engelhardt2011}. The resulting localization error can be reduced using polarization analysis~\cite{Backlund2012,Lew2014,Backlund2016} or dedicated PSF fitting~\cite{Enderlein2006,mortensen2010optimized,quirin2012optimal,stallinga2012position}, and vanishes for a focused image. Localization errors of comparable magnitude can occur when the emission pattern is distorted by near-field coupling to a nanoantenna \cite{Wertz2015plasmon, Raab2017shifting}. 

Here we show that methods for position estimation of emitters can be subject to large fundamental systematic errors when imaging elliptically polarized emitters as a consequence of spin-orbit coupling in the emitted light field. These errors are present even for ideal, focussed, aberration-free imaging systems.
Imaging a single trapped atomic ion as well as a single gold nanoparticle that emits light with different elliptical polarizations, we demonstrate a wavelength-scale shift between the measured and actual positions of the emitter. For a wide range of polarizations, this shift is nearly independent of the numerical aperture. However, it can become arbitrarily large for certain polarizations and vanishing numerical aperture. These findings reveal that, even for small numerical apertures, the paraxial approximation is fundamentally inadequate in the context of the centroid estimation method.

\begin{figure}
\centerline{\includegraphics[width=1\columnwidth]{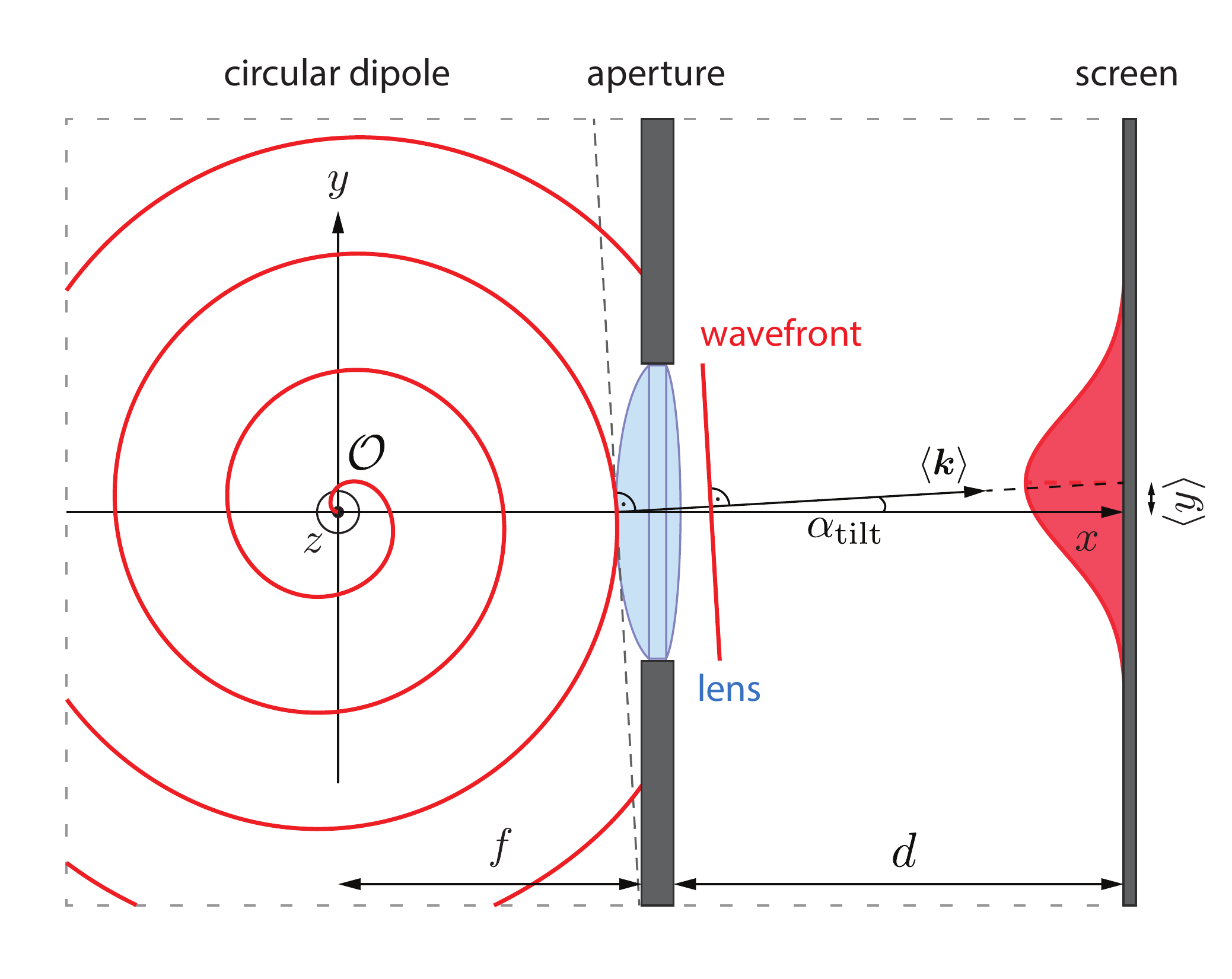}}
\caption{\textbf{Polarization-dependent displacement.} A $\sigma^{+}$ rotating dipole located at $\mathcal{O}$ emits spiral wavefronts in the equatorial $x$--$y$ plane, which are collimated by a lens with focal length $f$ centred on the $x$ axis and focused on the dipole. The wavefronts passing through the aperture of the lens have a mean wavevector $\langle \boldsymbol{k} \rangle$ tilted by an angle $\alpha_{\text{tilt}}$ with respect to the $x$ axis, which shifts the intensity distribution by $\langle y \rangle$ after a propagation length $d$. This shift originates from an orbital angular momentum of $\hbar$ per photon and results in an apparent displacement $\Delta y = -\lambda/2\pi$ of the emitter (see text). For a $\sigma^{-}$ emission the shift occurs in the opposite direction, since the wavefronts spiral in the opposite way.
}
\label{geometry}
\end{figure}

In order to understand the physical origin of the image shift, let us consider a circularly polarized dipole emitter rotating in the $x$--$y$ plane, at the centre $\mathcal{O}$ of the coordinate system. In this case, the total angular momentum carried by an emitted photon with respect to $\mathcal{O}$ is $\pm \hbar \bm{e_z}$, where $\pm$ corresponds to right-handed ($\sigma^+$-) or left-handed ($\sigma^-$-) polarization of the dipole relative to the $z$ axis, respectively. This total angular momentum can be decomposed into spin and orbital angular momentum, represented by the operators $\hat{S}_z$ and $\hat{L}_z$, respectively. The spin and angular momentum components of the dipole field are coupled and their expectation values for a $\sigma^\pm$-polarized dipole are
\begin{eqnarray}
\langle \hat{S}_z \rangle = \pm \hbar \frac{2\cos^2\theta}{1+\cos^2\theta}, \quad 
\langle \hat{L}_z \rangle = \pm \hbar \frac{\sin^2\theta}{1+\cos^2\theta}~,
\end{eqnarray}
where $\theta$ is the angle between the $z$ axis and the direction of observation~\cite{moeconservation,schwartz2006}. In the $x$--$y$ plane $(\theta=90^\circ)$, the photons carry exclusively orbital angular momentum with expectation value $\pm \hbar$ while the spin angular momentum vanishes, corresponding to linear polarization. This is an example of spin-orbit coupling of light~\cite{Bliokh2015} which gives rise to intriguing phenomena such as spin-Hall effect of light~\cite{Bliokh2008,Herrera2010} and chiral interactions between light and matter~\cite{chiral2016}.
For the circularly polarized dipole field, orbital angular momentum manifests as spiral wavefronts in the $x$--$y$ plane (Fig.~\ref{geometry}). Hence, the local wavevectors are tilted with respect to the radial direction and the linear momentum per photon has an azimuthal component with expectation value $\langle \hat{p}_\phi(r)\rangle=\langle \hat{L}_z \rangle/r =\pm \hbar / r$. Due to this tilt, the photons seem to originate from a position that is offset from the emitter~\cite{schwartz2006,li2010macroscopic}, a fact already predicted by Charles G. Darwin more than 80 years ago \cite{darwin1932notes}.

\begin{figure}
\centerline{\includegraphics[width=1\columnwidth]{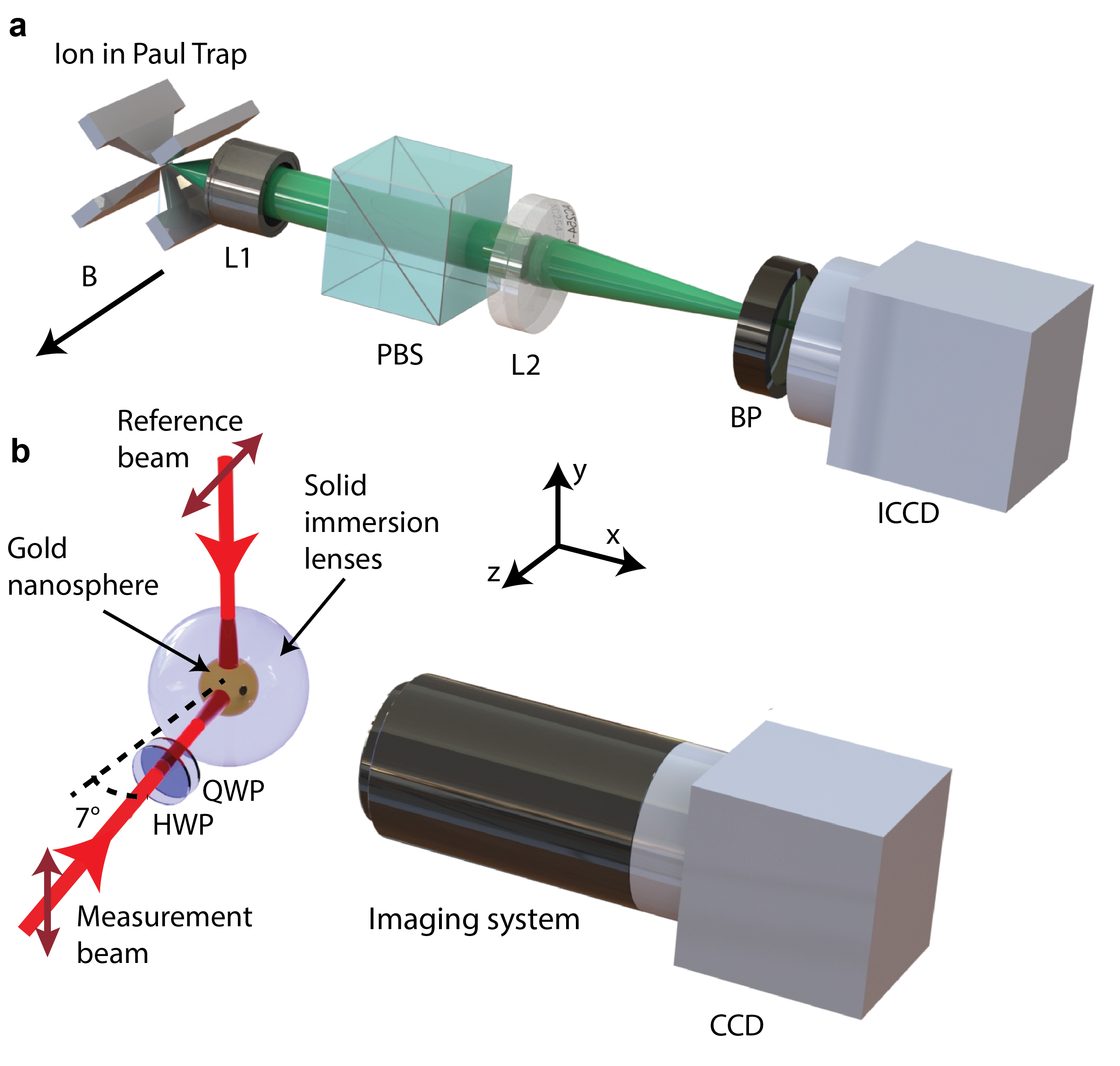}}
\caption{\textbf{Experimental set-ups.} \textbf{a,} A $^{138}$Ba$^+$ ion is confined in a linear Paul trap. A magnetic field $\textbf{B}$ along $\hat z$ defines the quantization axis and the rotation axis of the dipoles. Fluorescence light is collected in the $\hat x$ direction by an in-vacuum objective (L1, focal length: 25 mm, NA=0.40), and a lens (L2, focal length 150 mm) forms a focus on an intensified CCD camera (ICCD). A polarization beam splitter (PBS) filters out photons with polarization parallel to the quantization axis ($\pi$-polarized photons), while a bandpass filter (BP) selects photons with wavelength $493\pm1\,$nm. \textbf{b,} A gold nanosphere is located in the gap between two solid immersion lenses, filled with index matching oil to prevent reflections. The particle scatters light alternatively from a reference beam with fixed linear polarization and a measurement beam whose polarization is adjusted using half- (HWP) and quarter-wave (QWP) plates. The scattered light is collected by a microscope objective and imaged onto a CCD camera.
}
\label{setup}
\end{figure}

To quantify this shift for a typical imaging system, we consider a circularly polarized dipole emitter located at the front focal point of a lens with focal length $f$, centred on the $x$ axis. The lens collimates the light and changes its wavevector distribution. However, the mean wavevector $\langle \boldsymbol{k} \rangle$ averaged over the aperture is conserved and the collimated light propagates at an angle 
\begin{equation}
\alpha_{\rm tilt}= \frac{\langle \hat{p}_\phi\rangle_A}{\hbar k} \simeq \pm \frac{\lambda}{2\pi f}
\end{equation}
with respect to the optical axis. Here, $\langle \cdot \rangle_A$ denotes the expectation value per photon within the aperture $A$ of the lens. The centroid of the intensity distribution at a screen placed at a distance $d$ behind the lens is shifted in the $y$ direction by $\langle y \rangle = \alpha_{\rm tilt}d$ (Fig.~\ref{geometry}) and the apparent $y$ position of the dipole in the object plane is shifted by 
\begin{equation}
\Delta y=-\frac{f}{d}\langle y \rangle =\mp \frac{\lambda}{2\pi}~.
\end{equation}
This expression holds for any imaging system, replacing $f/d$ by the magnification factor of the system.
To summarize, due to spin-orbit interaction, the light emitted by a circularly polarized $\sigma^\pm$ dipole carries orbital angular momentum. When imaging in the plane of polarization of the dipole, this gives rise to a $\mp \lambda/(2\pi)$ shift of the apparent position of the emitter.

We now generalize the above for an elliptically polarized emitter oscillating in the $x$--$y$ plane. Its polarization state can be written as a superposition of $\sigma^+$- and $\sigma^-$-polarizations $|\psi\rangle=\alpha|\sigma^+\rangle+\beta|\sigma^-\rangle$, with $|\alpha|^2+|\beta|^2=1$. For a small numerical aperture $\text{NA}=D/(2f)\ll 1$, the shift of the apparent position of the emitter is (see Methods)
\begin{equation}
\Delta y=-\frac{\lambda}{2\pi}\cdot\frac{\Re(\epsilon)}{1+\text{NA}^2|\epsilon|^2/2}\label{eq:Deltay}
\end{equation}
where the \emph{dipole polarization ratio}, $\epsilon=(\alpha+\beta)/(\alpha-\beta)$, is in general complex and $\Re(\cdot)$ denotes the real value. For $\sigma^+$-polarization ($\sigma^-$-polarization) $\epsilon=+ 1$ ($\epsilon=- 1$) and for linear polarization along the $y$ axis ($x$ axis) $\epsilon=0$ $(\epsilon=\infty)$. For circular polarization and $\text{NA}\ll 1$ we recover the $\mp \lambda/(2\pi)$ shift derived above. When the axes of the polarization ellipse coincide with the $x$ and $y$ axes, $\epsilon$ is real and the shift is given by 
\begin{align}
\Delta y \simeq -\epsilon\frac{\lambda}{2\pi},\label{eq:linshift} 
\end{align}
as long as $|\epsilon|\ll1/\text{NA}$. Outside of this linear regime, the shift reaches a maximum $\Delta y_{\rm max}=\mp\lambda/(\sqrt{8}\pi\text{NA})$ for $\epsilon=\pm \sqrt{2}/\text{NA}$. Remarkably, this implies that the shift of the apparent position of the emitter can take arbitrarily large positive and negative values for small numerical apertures. For example, with $\text{NA}=0.23$, the distance between the two extremal shifts is as large as the optical wavelength $\lambda$. These large shifts are reached for $\epsilon = \pm 6.3$, \emph{i.e.}, when the polarization of the dipole is almost linear along the optical axis of the imaging system. In this case, the corresponding expectation values of the local orbital angular momentum per photon at the aperture significantly exceed $\hbar$, the total angular momentum per emitted photon. Such `supermomentum' \cite{bekshaev2015} is an example of weak value amplification common to structured optical fields, in which the local expectation value of an operator can take values outside its spectrum where the field is weak ~\cite{berry2009, knee2016weak}. We note that there is a close connection between the observed weak value amplification and the appearance of momentum vortices in the emitted light field. This connection is shown in Extended Data Fig. \ref{many_images} which plots the field distribution of the emitted light for different polarization states of the emitter. The plots also provide a graphical illustration for the polarization ratio $\epsilon$ which yields the maximum shift of the apparent position: This maximum shift is reached once the momentum vortices enter the field collected by the imaging lens. The centroid determination can be interpreted as a measurement of the weak value of the photons' orbital angular momentum (see Methods). Finally, we note that the predicted shifts also occur for large numerical apertures and that Eq.~(\ref{eq:linshift}) remains approximately valid provided that $|\epsilon|\lesssim 1$ (see Methods).

We study the predicted shifts by imaging a single atom --- a fundamental quantum emitter --- and a single sub-wavelength scale nanoparticle. In the first experiment, we confine a $^{138}$Ba$^+$ atomic ion in a Paul trap and image fluorescence from the dipole transition at $\lambda_1 = 493.41\,$nm (Fig.~\ref{setup}a) using an imaging system with magnification $M_a = 5.40(7)$ and $\text{NA} = 0.40$ (see Methods). A bandpass filter and a polarizer are used to collect light selectively from one of the spontaneous decay channels of the excited state, corresponding to the emission from either a $\sigma^+$ or a $\sigma^-$ dipole (see Methods). 

\begin{figure*}
\centerline{\includegraphics[width=0.98\textwidth]{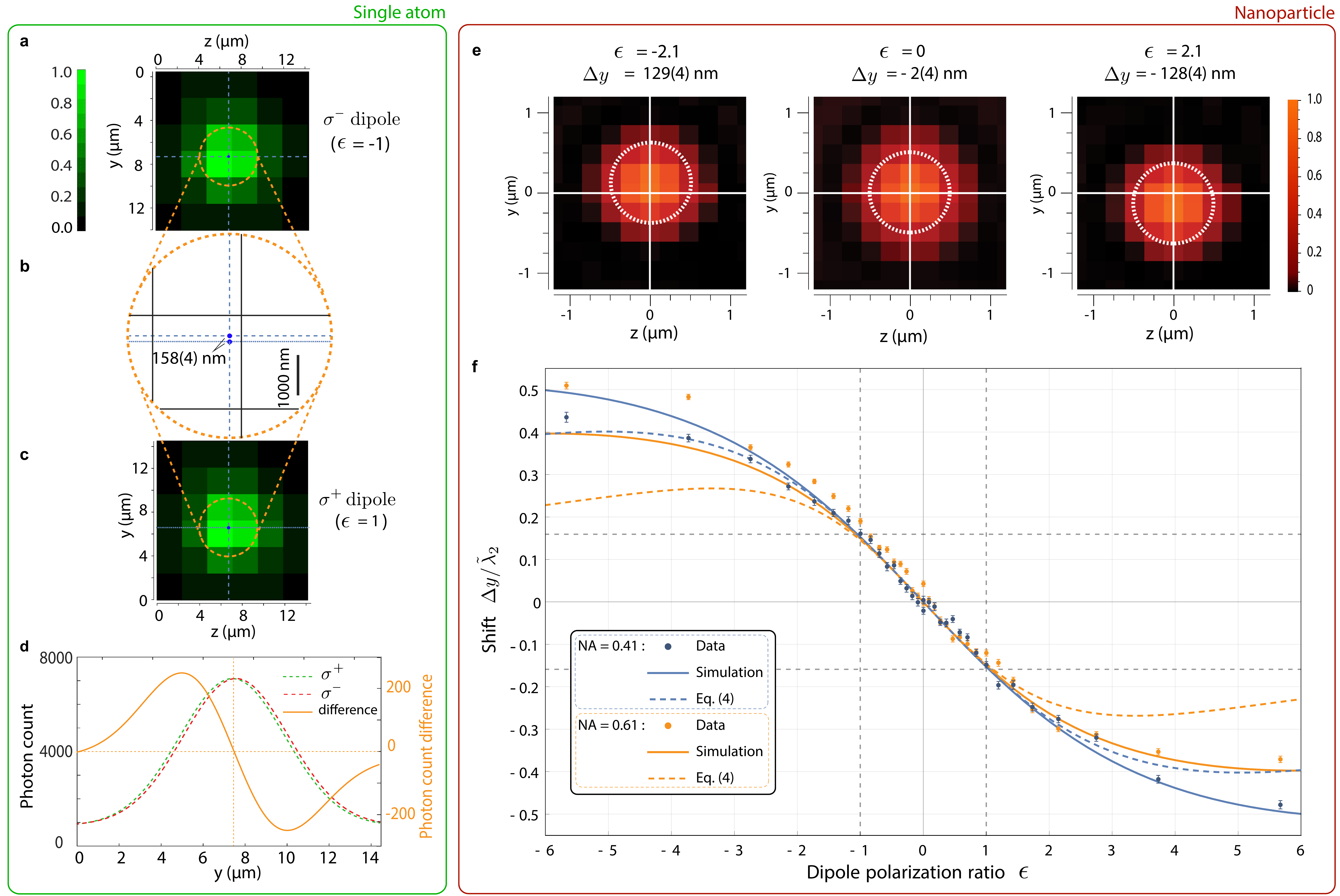}}
\caption{\textbf{Apparent displacement of the emitters.} \textbf{a,} \textbf{c,} Measured images (normalized to maximum pixel count rate) of a single atom for the $\sigma^-$ and $\sigma^+$ transitions. The blue lines and blue points indicate the centroid of the image obtained by a 2D Gaussian fit to the data. The orange circle represents the 1$\sigma$-width. \textbf{b,} Zoom of the centre of images \textbf{a} and \textbf{c}. The two blue points show the centroid position of the fitted $\sigma^-$ (upper point) and $\sigma^+$ (lower point) images. \textbf{d,} Vertical cross section of the Gaussian fits for $\sigma^+$ (green dashed curve, left scale) and $\sigma^-$ (red dashed curve, left scale) polarizations. The orange curve shows the difference of both fits (right scale). \textbf{e,} Measured images of the nanoparticle for $\epsilon=\pm 2.1$ and $\epsilon=0$ for $\text{NA}=0.41$. The white cross indicates the position of the nanoparticle obtained from the reference image. The dashed circle with a diameter of $500$ nm indicates the $1\sigma$ width of the image obtained from a Gaussian fit and is centred around the apparent position of the nanoparticle. \textbf{f,} Relative displacement of the image of the particle as a function of $\epsilon$, measured for two different NAs. The error bars indicate the $1\sigma$ statistical error. The dashed curves are the theoretical predictions of Eq.~(\ref{eq:Deltay}) and the solid curves are the displacements obtained by simulations of the image process taking into account that the centroid of the images are obtained from a Gaussian fit (see Methods). The dashed grey lines show the case of circularly polarized emitters.}
\label{results_image}
\end{figure*}

We estimate the emitter's position from each image by fitting a 2D Gaussian function, which is a suitable approximation to the PSF in the measured regime \cite{stallinga2010} (see Methods). Fig.~\ref{results_image}a-c show the results for a total measurement time of 3 hours. We observe a displacement between the $\sigma^+$ and $\sigma^-$ emissions of 158(4)\,nm in the object plane, in agreement with the expected value $\lambda_1/\pi = 157.1\,$nm.

As it is demanding to generate an arbitrarily polarized emission from a single atom, we extend the study to the case of a general elliptical polarization in a separate experiment where we image the light scattered by a single sub-wavelength-sized spherical gold nanoparticle. Such particles are used as labelling agents for super-resolution microscopy in biological research~\cite{howes2014colloidal,zhang2015super}. Being a spherically symmetric emitter, the polarization of a nanoparticle's dipole always coincides with the polarization of the illuminating field, which can be controlled precisely. We place a 100\,nm-diameter gold nanoparticle in the centre of a glass sphere with refractive index $n = 1.46$ by depositing it on an optical nanofibre~\cite{petersen2014chiral} and surrounding it by two fused silica 2.5\,mm-radius hemispherical solid immersion lenses. The $\sim200\,\mu$m gap between the lenses is filled with index matching oil to prevent any reflection near the particle from either the nanofibre or the lenses. The nanoparticle is illuminated by a laser beam (vacuum wavelength $\lambda_2=685\,$nm) with adjustable polarization and the scattered light is imaged onto a CCD camera through the sphere and a microscope (Fig.~\ref{setup}b). To test the dependence of the position shift on the NA, two different microscope objectives are used with the same nominal magnification but different numerical apertures, resulting in $\text{NA} = 0.41$ and $\text{NA} = 0.61$ when including the silica sphere, and magnifications $21.9(2)$ and $20.1(1)$, respectively. The apparent displacement of the nanoparticle is measured by fitting a 2D Gaussian function to its image (see Methods), using alternatively the beam with adjustable polarization and a linearly polarized reference beam. The measurements, averaged over 125 individual realisations for each polarization setting, are shown in Fig.~\ref{results_image}.
For $|\epsilon|<2$, within our experimental errors, we observe a very good agreement of our measurements with the expected linear increase of the displacement with $\epsilon$, independent of the numerical aperture. For larger $|\epsilon|$, the linear approximation is not valid and the experimental data follows approximately the theoretical prediction from Eq.~(\ref{eq:Deltay}) (dashed lines). The apparent positions of the nanoparticle imaged with right and left circular polarizations ($\epsilon = \pm 1$) are displaced relative to each other by $145(6)\,$nm for $\text{NA}=0.41$ and $146(4)\,$nm for $\text{NA}=0.61$, in agreement with the expected value $2\Delta y = \tilde\lambda_2/\pi \approx 150\,$nm, where $\tilde\lambda_2 = \lambda_2/n$ is the laser wavelength in the index matching oil. The displacement increases for larger values of $|\epsilon|$, and the total displacement between counter-rotating elliptical polarizations reaches $430(7)\,$nm ($\simeq\tilde\lambda_2$) for $\epsilon = \pm 5.67$, a shift four times larger than the diameter of the gold nanoparticle. In order to verify that focusing errors are not at the origin of the effect, we slightly defocus our imaging optics and observe that, in the measured range, the shifts do not depend on the distance of the particle to the focal plane (see Methods).

Our findings may affect super-resolution microscopy techniques, which achieve resolutions two orders of magnitude smaller than the systematic shifts demonstrated here~\cite{Yildiz2003,small2014superresolution}. For instance, the determination of the position of an emitter with $\text{NA}=1$, at a wavelength of $\lambda\approx 628\,$nm with an accuracy of 1\,nm, requires the scattered light to be more than $99.99~\%$ linearly polarized ($|\Re(\epsilon)|<0.01$, see Methods). For larger $\epsilon$, an accuracy of, \emph{e.g.}, 1 nm could still be reached by employing an algorithm that not only uses position but also polarization of the dipole as fit parameters for the recorded point-spread function. However, in order to reach the necessary signal-to-noise ratio, this higher dimensional fit requires one to increase the light-collection time by more than 4 orders of magnitude compared to the case of an optimally coupled linear dipole (see Methods and Extended Data Fig.~\ref{Precision_limit_Fig}d). The residual contributions of elliptical polarizations that give rise to the presented effect are in general difficult to avoid in realistic situations, and therefore, the discussed systematic fundamental error is always present (see Methods).

On the positive side, the polarization-dependent shift could be used, \emph{e.g.}, in arrays of optically trapped particles~\cite{Bakr2009}, where the apparent location of each particle would give access to the local polarization of an inhomogeneous exciting field; conversely, in the case of an homogeneous exciting field, the shift would allow to sense local physical parameters affecting the polarizability of the particles, such as the direction of the magnetic field. The demonstrated effect is relevant beyond optical imaging, as it will occur for any kind of wave carrying transverse orbital angular momentum. Thus, it may affect the localization of remote objects imaged with radar or sonar techniques~\cite{lee2009polarimetric,Hayes2009SAS}, or even alter the apparent position of astronomical objects detected through their emission of gravitational waves \cite{abbott2017search,bialynicki2016gravitational}.

\bibliographystyle{naturemag}
\bibliography{library}

\section*{Acknowledgements}
\noindent We thank P. Ob\v{s}il for his experimental support, and J. Enderlein, M. Hush and A. Jesacher for helpful discussions. This work has been supported by the Austrian Science Fund (FWF, SINPHONIA project P23022, SFB FoQuS F4001, SFB NextLite F4908), by the European Research Council through project CRYTERION \#227959 and by the Institut f\"ur Quanteninformation GmbH.

\section*{Author contributions}
\noindent J.V. and A.R. proposed the concept. All authors contributed to the design and the setting-up of the experiments (atom experiment: G.A., Y.C., D.B.H., and R.B.; nanoparticle experiment: S.W., J.V., and A.R.). G.A. and D.B.H. performed the atom experiment and analysed the data. S.W. performed the nanoparticle experiment and analysed the data. All authors contributed to the writing of the manuscript. 

\section*{Additional information}
\noindent Correspondence and requests for materials should be addressed to G.A., A.R., or J.V.

\section*{Competing financial interests}
\noindent The authors declare no competing financial interests.


\clearpage

\renewcommand{\figurename}{Methods Figure}
\setcounter{figure}{0} 
\setcounter{equation}{0}
\setcounter{page}{1}
\renewcommand\thesubsection{\arabic{subsection}}

\thispagestyle{empty}
\onecolumngrid
\begin{center}
\vspace{5 mm}
\textbf{\large Methods for:\\Wavelength-scale errors in optical localization due to spin-orbit coupling of light}\\
\vspace{2 mm}
G. Araneda,$^1$ S. Walser,$^{2}$ Y. Colombe,$^1$ D. B. Higginbottom,$^{1,3}$ J. Volz,$^{2}$	R. Blatt,$^{1,4}$ and A. Rauschenbeutel$^{2}$\\
\vspace{2 mm}

\textit{\small
$^1$Institut f\"{u}r Experimentalphysik, Universit\"{a}t Innsbruck, Technikerstra\ss e 25, 6020 Innsbruck, Austria\\
$^2$Vienna Center for Quantum Science and Technology, TU Wien-Atominstitut, Stadionallee 2, 1020 Vienna, Austria\\
$^3$Centre for Quantum Computation and Communication Technology, Research School of Physics and Engineering, The Australian National University, Canberra ACT 2601, Australia\\
$^4$Institut f\"{u}r Quantenoptik und Quanteninformation, \"{O}sterreichische Akademie der Wissenschaften, Technikerstra\ss e 21a, 6020 Innsbruck, Austria\\
}

\vspace{10 mm}
\end{center}
\twocolumngrid

\section{Calculation of the centroid position in optical imaging}
\subsection*{Wavefronts of the radiated field}
The electric field emitted by an optical dipole located at the origin ($r=0$) that oscillates with angular frequency $\omega$ is given by
\begin{equation}
\boldsymbol \Psi(\boldsymbol r,t) = -\frac{\omega^2}{4\pi\epsilon_0c^2} \frac{e^{i( k r-\omega t)}}{r^3} (\boldsymbol{r}\times\boldsymbol{\mu})\times \boldsymbol{r} \label{eqn:Efield}
\end{equation}
in the far field ($|\boldsymbol{r}|\gg\lambda$), where $\boldsymbol{\mu}=\mu\boldsymbol{e}_\mu$ is the complex vector amplitude of the electrical dipole and $k=2\pi/\lambda$ where $\lambda$ is the wavelength of the emitted light.
From Eq.~(\ref{eqn:Efield}), it is possible to derive an expression for the wavefronts, \emph{i.e.} the surfaces of constant phase of the electromagnetic wave. For a linearly polarized dipole (a dipole with zero expectation value for its angular momentum) the wavefronts are spheres given by $r_{wf}=\frac{\omega t}{k}+\text{const}$, whereas for a $\sigma^+$ or $\sigma^-$ polarized dipole that radiates waves with total angular momentum per photon of $\pm\hbar$ with respect to the $z$ axis, the wavefronts in the $x$--$y$ plane are given by the parametric equation
\begin{eqnarray}
	r_\pm(\phi)&=&\frac{\mp\phi+\omega t}{k}+\text{const}. \label{eqn_spiral}
\end{eqnarray}
This corresponds to an Archimedean spiral rotating around the $z$ axis, with the same rotation sense as the dipole.

\subsection*{Angular momentum and imaging}
For any point in space, we can assign a local orbital angular momentum to the light field, which can be calculated by applying the operator 
\begin{equation}
\hat{\boldsymbol L}= \boldsymbol r \times\hat{\boldsymbol p}, \label{eqn:L}
\end{equation} 

on the wave function, where $\hat{\boldsymbol p}=- i\hbar \vec{\nabla}$ is the momentum density operator. The local orbital angular momentum per photon can be measured by sending the light through an aperture at position $\boldsymbol r_0$. Measuring the displacement $\langle q\rangle$ of the centre of mass of the far-field image from the optical axis $\boldsymbol e_r$ at distance $d$ from the aperture gives the expectation values of the transverse linear momentum components $\langle\hat p^w_q\rangle$ per photon at the position of the aperture, where $q=(x,y,z)$ are the Cartesian coordinates with respect to the dipole at the origin. The relation between angular momentum and displacement is given by
\begin{equation}
\langle q\rangle = \frac{d}{\hbar k} \langle \hat p^w_{q}\rangle = \frac{d}{\hbar k} \frac{1}{r_0} \langle \hat L^w_{q}\rangle.
\end{equation}
This measurement can be interpreted in the framework of weak measurements, where the centre of mass in the image plane is proportional to the weak value of the photons' orbital angular momentum (or the transverse linear momentum) at the aperture, which are given at the position of the first lens by \citeMethods{berry2009M} 
\begin{eqnarray}
\langle \hat L_{q}^w\rangle=r_0 \cdot\langle \hat p^w_{q}\rangle=r_0\cdot\ \frac{\langle\boldsymbol{\tilde\Psi_{\text{post}}}|\hat p_{q}|\boldsymbol\Psi\rangle} {\langle\boldsymbol{\tilde\Psi_\text{post}}|\boldsymbol\Psi\rangle} \,, \label{eqn:weak}
\end{eqnarray}
where $\boldsymbol{\tilde \Psi_\text{post}}$ is the part of the wavefunction that passes the aperture (the post-selected state).

In other words, the orbital angular momentum components transverse to the optical axis result in a transverse linear momentum at the aperture that leads in turn to a displacement of the diffracted beam in the far field. The local angular momentum per photon can exceed $\hbar$ where the field is weak (so-called `supermomentum') \citeMethods{bekshaev2015M}. The centre of mass of the far-field image can be considered as a measurement of the weak value of the angular momentum of the light that passes through the aperture.

In realistic imaging systems, a lens is included at the position of the aperture such that the emitter is in the focal plane at distance $f$ from the aperture. An ideal lens applies a phase transformation to light that passes the aperture, such that any wave originating from a single point in the focal plane is transformed into a plane wave. For any such wave, the average wavevector and thus the average transverse momentum is conserved. 

\subsection*{Calculation of the image centroid}
For the derivation of the displacement of the centroid of the image in a typical imaging system, we start with the above relation between angular momentum and transverse linear momentum when our objective has an aperture with diameter $D$. We consider the situation where the angular momentum of the light is fully transverse to the optical axis of the imaging system ($x$ axis) and we set our quantization axis ($z$ axis) along the angular momentum direction. The imaging system consists of an objective with focal length $f$ located at a distance $f$ from the emitter. In this situation, the electric fields of the three elementary dipoles $\pi$, $\sigma^+$ and $\sigma^-$ at the objective are, for small aperture ($D\ll f$), given by
\begin{eqnarray}
\tilde{\boldsymbol\Psi}_\pi(\rho,\phi)&=&\frac{1}{f}\boldsymbol{e_z}e^{i\varphi}, \label{eqn:PsiPi}\\
\tilde{\boldsymbol\Psi}_{\sigma^\pm}(\rho,\phi)&=&\frac{1}{\sqrt{2}}\left(\pm\frac{i}{f}\boldsymbol{e_y}+\frac{\rho}{f^2}\boldsymbol{e_\rho}\right)e^{i\varphi},\label{eqn:PsiSigma}
\end{eqnarray}

where $\rho$ and $\phi$ ($y$ and $z$) are polar (Cartesian) coordinates in the aperture plane, $\boldsymbol{e_x}$, $\boldsymbol{e_y}$, $\boldsymbol{e_z}$ and $\boldsymbol{e_\rho}$ are the unit vectors in the respective direction, $\varphi=k\sqrt{\rho^2+f^2}$. $\tilde\Psi_{\pi,\sigma^{\pm}}$ are the parts of the wavepacket that pass through the aperture from the corresponding dipoles. Since the emitter is in the focal plane of the objective, the latter applies the transformation $e^{-i\varphi}$ on the light and removes the phase factor in Eqs.~(\ref{eqn:PsiPi}) and (\ref{eqn:PsiSigma}) which we drop in the following. In the case where the light has no orbital angular momentum, the transformation of the objective results in a beam with planar wavefronts perpendicular to the optical axis. Consequently, the light has no linear momentum transverse to the optical axis. For the case where the incoming light has orbital angular momentum along the $z$ axis, the wavefronts after the objective are tilted with respect to the optical axis and the light has linear momentum in a direction transverse to the optical axis. Measuring the displacement of the waveform's centre of mass from the optical axis $\langle q \rangle$ at distance $d$ from the objective ($d \gg D$) then corresponds to a measurement of the expectation value of the transverse angular momentum component per photon $\langle\hat L^w_{q}\rangle$ or the linear transverse momentum component $\langle\hat p^w_{q}\rangle$ of the photons at the position of the aperture where $q \in (y,z)$. The actions of the momentum operators on the wave are
\begin{eqnarray}
\hat p_{q}\tilde{\Psi}_{\sigma^\pm} &=&\pm\frac{i\hbar}{f^2\sqrt{2}}\boldsymbol{e}_{q},
\end{eqnarray}
as well as $\hat p_{q} \tilde{\Psi}_\pi=0$. Considering the general case of a photon that originates from a superposition of $\sigma^+$ and $\sigma^-$ emission, \emph{i.e.}, $\Psi=\alpha\Psi_{\sigma^+}+\beta\Psi_{\sigma^-}$, we can calculate the weak value in Eq.~(\ref{eqn:weak}) and obtain 
\begin{eqnarray}
\langle\hat p^w_y\rangle&=&\frac{\hbar}{f}\frac{\Re(\epsilon)}{1+|\epsilon|^2 \text{NA}^2/2}\label{eqn:displace},\\
\langle \hat p^w_z\rangle&=&0.
\end{eqnarray}
Here, we defined the numerical aperture $\text{NA}=D/(2f)$ and the complex valued amplitude ratio or dipole polarization ratio $\epsilon=(\alpha+\beta)/(\alpha-\beta)$ of the two polarization components, with $|\alpha|^2+|\beta|^2=1$. Therefore, the wavepacket only exhibits a transverse displacement along the axis in the imaging plane which is perpendicular to the axis of its angular momentum.

\subsection*{Microscopy set-up}
In a microscopy set-up, the image is not formed at infinity, but a second lens with focal length $f'$, which we assume to be at a distance $f'$ from the aperture, is used to form an image at position $x=2f'$. In this case, the expected displacement is obtained by replacing $d$ by $f'$. This finally yields for the expected displacement on the screen
\begin{eqnarray}
\langle\hat y\rangle &=&\frac{1}{\hbar k}\frac{f'}{f}\langle \hat L^w_{y}\rangle=\frac{\lambda}{2\pi}\frac{f'}{f}\frac{\Re(\epsilon)}{1+|\epsilon|^2 \text{NA}^2/2}.\label{eqn_centroid}
\end{eqnarray}
Eq.~(\ref{eqn:displace}) has two noteworthy consequences. First, for small numerical aperture ($\text{NA}\ll|\epsilon|$) and $\epsilon$ real, the displacement of the centroid increases linearly in $\epsilon$. Second, in the case of circular polarization $\epsilon=\pm1$, the centroid of the image is displaced from the expected position by $\langle\hat y\rangle\approx\pm\lambda/(2\pi)$ times the magnification of the optical system $f'/f$, \emph{i.e.}, the particle appears to be displaced by $\lambda/(2\pi)$. The maximum displacement of the centroid for $\epsilon$ real is given by
\begin{equation}
\langle\hat y\rangle_{\rm max} =\pm\frac{\lambda}{2\pi}\frac{f'}{f}\frac{1}{2\text{NA}},\label{eqn_centroid:max}
\end{equation}
\emph{i.e.}, for vanishing numerical aperture, the displacement of the apparent and real positions of the particle can be arbitrarily large. Similarly, the local momentum per photon is known to diverge around optical vortices \citeMethods{barnett2013M}, in speckle patterns \citeMethods{dennis08}, and in the interference of two plane waves \citeMethods{berry2009M,bekshaev2015M} (which $\Psi$ approaches in the small-aperture limit).

\subsection*{Fourier-optic derivation of the centroid position}
We note that the position of the centroid can also be calculated in the framework of Fourier-optics. We can calculate the electric fields of the three fundamental electrical dipoles oscillating in $x$, $y$ and $z$ directions in the image plane and obtain for the approximation of small $\text{NA}$
\begin{eqnarray}
\boldsymbol{E}_x&=&-iE_0\cdot \frac{\text{NA}^2}{\rho} J_2(\tilde{\rho}) (\cos\varphi\boldsymbol{e_y}+\sin\varphi\boldsymbol{e_z}),\label{eq_image1} \\
\boldsymbol{E}_y&=&E_0\cdot \frac{\text{NA}}{\rho} J_1(\tilde{\rho})\boldsymbol{e}_y,\label{eq_image2}\\
\boldsymbol{E}_z&=&E_0\cdot \frac{\text{NA}}{\rho} J_1(\tilde{\rho})\boldsymbol{e}_z,\label{eq_image3}
\end{eqnarray}
where we have defined the amplitude 
\begin{equation}
E_0=\frac{\mu\omega^2}{4\pi\epsilon_0^2c^2},
\end{equation} 
and $\tilde{\rho}=\rho\cdot k \cdot \text{NA} \cdot f/f'$, with the opening angle of the objective $\text{NA}\approx D/(2f)$. The final image is then a superposition of the three dipole fields from which we obtain for the centroid again Eq.~(\ref{eqn_centroid}).

It is also possible to numerically calculate the intensity distribution in the image plane by a full propagation of the electromagnetic fields through the optical system~\citeMethods{novotny2012principles}. This allows one to calculate the images for arbitrarily large NA. Extended Data Fig.~\ref{many_images} and \ref{displacements} show the images and centre of mass positions calculated in this way for different values of $\epsilon$ and $\text{NA}$ , respectively.

\subsection*{Displacement in immersion microscopy}
\label{immersion_microscopy}
All the above considerations originate from the orbital angular momentum of the light, which is linked to the curvature of its wavefronts in the far field. In high-NA imaging, the so-called immersion method is used, where the first lens of the system is a solid immersion lens, \emph{i.e.}, a half-ball lens. In this method, the imaged particles are located on the planar side of the lens and embedded in immersion fluid that has the same refractive index as the lens. Consequently, wavefronts emitted by the particle are parallel to the surface of the lens and, thus, this method does not affect the wavefronts in the far field outside the lens. Therefore, the discussion presented above also applies for this case, where it is only necessary to replace the numerical aperture NA with the geometrical numerical aperture $\text{NA}_g$ ($\text{NA}_g=\text{NA}/n$), and for the calculated displacement replace $\lambda$ with the wavelength in the immersion fluid $\lambda/n$.

\subsection*{Backplane polarization filtering}
The orbital angular momentum of the light is associated with the emission of light from a circular dipole that can be decomposed into two linear dipoles, one oscillating parallel and one perpendicular to the optical axis. In the back focal plane the dipole parallel to the optical axis generates a radially polarized light field, while the polarization of the light from perpendicular dipole is mostly linear. Using an azimuthal polarization filter, \emph{i.e.}, a filter that blocks radial polarization, it is possible to block the light emitted from the longitudinal dipole. In this case, the final image is not shifted, but its shape is distorted. This method has already been demonstrated for the focus-dependent shifts and would also allow one to reduce the localization error due to the orbital angular momentum~\citeMethods{Lew2014M,Backlund2016M}.

\section{Implications for super-resolution microscopy}
\subsection*{Precision of position estimation}
In the following, we quantify how the discussed effect influences the achievable precision in the localization when the polarization of the emitter is unknown. For this purpose, we compare the achievable localization precision in such an algorithm when considering linearly polarized emitters only, and when more general elliptical emitters are considered. In real situations, the signal-to-noise ratio of the acquired images prevents that all the parameters of the emitting dipole (position and dipole polarization) can be identified with arbitrary precision. As a consequence, for limited signal-to-noise ratio, it is not possible to distinguish if an image originated from a displaced linearly polarized dipole or from an non-displaced elliptical dipole which causes an apparent shift.

In order to get an estimate of the error that the presence of elliptical dipoles introduces into these methods, we assume the following situation: a linear emitter, with dipole orientation along the $y$-axis (orthogonal to the detection axis) is located at the origin of the coordinate system and imaged onto a CCD-chip. The optical axis of the imaging system equals the $x$-axis. In the case of infinite signal-to-noise, the point-spread function is given by Eq.~(\ref{eq_image2}). For a given pixel-size this yields the discrete photon number distribution $n_i^{\text{lin}}$ on the CCD, where $i$ indicates a given pixel. We then analyse how well this point-spread function fits to the one of an elliptically polarized dipole emitter, with dipole polarization ratio $\epsilon$ as defined in the main text, and located at a vertical distance $\delta y$ from the origin. According to our findings, the image of this elliptical dipole is additionally shifted by $M\Delta y(\epsilon)$, so that the centroid of the image is located at $M\cdot(\delta y+\Delta y(\epsilon))$, and the image at the CCD-camera is then given by the discrete distribution $n_i(\delta y,\epsilon)$. Here $M$ denotes the magnification of the imaging system. To compare the two distributions we define the normalized quadratic sum of the difference of the two images by
\begin{equation}
 S(\delta y,\epsilon)=\frac{1}{N^2}\sum_i (n_i^{\text{lin}}-n_i(\delta y,\epsilon))^2,
\label{eqn:1}
\end{equation}
where $N=\sum_i n_i^{\text{lin}}=\sum_i n_i(\delta y,\epsilon)$ is the total photon number in each image. This function is plotted in Extended Data Fig.~\ref{Precision_limit_Fig}a and b for the optimal ratio of pixel and point-spread function size. In the figure, an elongated minimum of $S(\delta y,\epsilon)$ is observed along the line $\delta y=\epsilon \cdot \lambda /(2\pi)$. This indicates that there exists a continuous set of correlated parameters $(\delta y,\epsilon)$ that can create an image almost indistinguishable from the one produced by a linear dipole. The apparent shift $\Delta y = -\epsilon \cdot \lambda / (2\pi)$ is in these cases cancelled by the real displacement $\delta y$. Still $S$ is at these points non-zero due to the slight distortion of the PSF because of the elliptical polarization.

Extended Data Fig.~\ref{Precision_limit_Fig}c shows $S(\delta y,\epsilon)$ evaluated along $\epsilon = \delta y \cdot 2\pi/\lambda$ (yellow points) and, for comparison, evaluated along $\epsilon=0$ (blue points). For small $\delta y$ and $\epsilon$, the function $S(\delta y,\epsilon)$ can be approximated by
\begin{eqnarray}
S(\delta y,\epsilon) &\approx& a\cdot \delta y^{2} \quad \text{for}\quad \epsilon = 0,\\
S(\delta y,\epsilon) &\approx& b\cdot \delta y^{4} \quad\text{for}\quad \epsilon = \delta y \cdot 2\pi/\lambda.
\end{eqnarray}
with $a\approx 33.3\cdot 10^{-3}$ and $b\approx 9.1\cdot 10^{-3}$.The approximated curves are shown in Extended Data Fig.~\ref{Precision_limit_Fig}c (solid lines).

The function $S(\delta y,\epsilon)$ describes the expectation value of the sum of mean squares for the ideal case of a noiseless image. In reality, however, each measured image contains noise. In the most fundamental case, this noise is purely photon shot-noise, \emph{i.e.}, the photon number per pixel $n_i$ fluctuates with the standard deviation $\Delta n_i=\sqrt{n_i}$. As a consequence, when considering a realistic image $n_i^{\text{exp}}$ of a linearly polarized dipole and fitting this image with the discretized point-spread function $n_i^{lin}$, we obtain a non-zero value for $S$ which changes from shot to shot. The expectation value of $S$ in this case is given by
\begin{equation}
\langle S_N\rangle=\frac{1}{N^2}\sum_i (n_i^{\text{exp}}-n_i^{lin})^2=\frac{1}{N}
\label{eqn:2}
\end{equation}
This expectation value $\langle S_N\rangle$ defines the accuracy of our position estimation. If we have no prior knowledge of the particle's polarization, fitting the experimental data will now yield a set of different tuples ($\delta y$,$\epsilon$) for each experimental run. The possible tuples, when considering shot-noise only, are those where $S(\delta y,\epsilon) \lesssim 1/N$ since in these cases the outcome of $S$ is on average dominated by the noise and not by $\delta y$ or $\epsilon$. This allows the direct calculation of a precision limit $Dy(N)$ which we define as the outer limit of the range of possible results $(\delta y, \epsilon)$ for a given number of collected photons $N$. This is realized by finding the parameters $Dy$ and $D\epsilon=Dy \cdot 2\pi / \lambda$ which yield $S(Dy,D\epsilon) = 1/N$. Similarly, the points $S(Dy,\epsilon=0) = 1/N$ give the precision limit in case of a solely linearly polarized emitter. 

Following this approach, Extended Data Fig.~\ref{Precision_limit_Fig}d shows the precision limit $Dy$ as a function of the number of detected photons $N$ and the numerical aperture of the imaging system for the case of a linearly (blue) and an elliptically polarized dipole (yellow). The upper $x$-axis corresponds to the residual uncertainty $D \epsilon$ of the emitter's polarization. The figure shows that the introduction of elliptical polarization decreases the achievable precision by orders of magnitude. The two precision limits (\emph{i.e.}, for linear and elliptical polarization) decrease with different power laws as a function of the number of photons. As a consequence, the ratio of the two errors diverges when increasing the localization precision.

For the particular algorithm used, and assuming a localization precision of 1 nm for typical parameters ($\text{NA}\sim 1$ and $\lambda\approx 628$ nm), the precision decreases by more than one order of magnitude and, thus, in order to get the same precision as in the case of a purely linear dipole ($\epsilon=0$) one has to increase the measurement time by more than 4 orders of magnitude (see Extended Data Fig.~\ref{Precision_limit_Fig}d grey dashed lines). Alternatively, one has to know the polarization of the emitter better than $D\epsilon=0.01$. This corresponds to an polarization overlap with a linear emitter of $\eta = (1+D\epsilon^2)^{-1}=99.99 \%$, which is a rather challenging requirement.

The presented analysis is based on a least-mean-square fit algorithm. While this might not be an optimal method, similar scaling of the errors should occur for more sophisticated algorithms, \emph{e.g.}, based on maximum likelihood estimation \citeMethods{mortensen2010optimizedM}.

\subsection*{Polarization-dependent shifts in real-life microscopy}
The effect demonstrated in this work relies on the emission of elliptically polarized light from individual emitters. While in our demonstration the emitters provided either a scalar polarizability (gold nanosphere) or transitions with circular eigenpolarizations (ion), emission of elliptically polarized light is in general expected for any type of physical emitter in many physical situations. 

In contrast to the unphysical model of a linearly polarized two-level emitter, a realistic emitter such as a fluorescent molecule always features different transitions with linearly independent transition moments and thus, the illuminating light will always couple to all possible polarization components. In principle, an isolated and unperturbed molecule can feature a linearly polarized transition that is singled out concerning its transition frequency and strength, such that the contributions of other transitions may be small because they are far-detuned. However, if the molecule is immersed in a gaseous, liquid, or solid medium, molecular collisions, temperature-induced vibrations, and/or strain-induced static distortions will shift the transitions and broaden them by several orders of magnitude. This then leads to a sizeable overlap between transitions that are spectrally well-separated in unperturbed case. Under these circumstances, the resulting complex-valued anisotropic polarizability of the molecule enables the emission of elliptically-polarized light even at resonant excitation of the singled-out linearly-polarized transition of the unperturbed molecule. 

In a realistic microscopy setting, two possible mechanisms will result in the emission of circular polarization and thus can lead to an apparent shift of the position of such an emitter. On the one hand, if the illumination light contains elliptical polarization components, the light scattered by the emitter can also contain elliptical polarizations. Even in the case where the emitter is freely rotating within the integration time, a residual elliptical polarization and thus an apparent shift will remain. We note that, for all practical purposes, elliptical polarizations in the illumination light will be present. Even for illumination with a perfectly linearly polarized light field, a spurious reflection of $4\%$ can lead to local ellipticities of up to $\epsilon=0.2$ which would lead to an apparent shift of about 20 nm for an emitter with scalar polarizability. Besides, elliptical polarizations arise when focussing an initially linearly polarized illumination beam \citeMethods{Bliokh2015b}. On the other hand, the dipole of an emitter with a complex-valued anisotropic polarizability can become elliptical even for perfectly linearly polarized illumination light and an apparent shift can occur. This situation is realized, \emph{e.g.}, for emitters with spectrally separated but overlapping linearly-polarized transitions such as plasmonic nanorods or fluorescent molecules with strongly broadened transitions. 

We note that even for cases where the elliptical polarization contribution are small, \emph{e.g.}, because the linearly polarized transitions of the molecules only weakly overlap, these effects cannot be neglected. Already a contribution of a circular polarization component on the order of $-40$ dB can lead to apparent shifts on the nanometer level, comparable or larger than the accuracy of state-of-the-art super-resolution microscopy systems. Furthermore, as we experimentally demonstrated, for relatively small NA, a weak elliptical polarization contribution is precisely the situation under which the apparent shifts can become particularly large and even spurious elliptical polarization component of the emitted light may lead to a wavelength-scale systematic shift of the image centroid.

\section{Experimental methods in the atom experiment}
\subsection*{Atomic transition selective detection of photons}
The photons are emitted from a dipole transition with angular momentum $\Delta m$ of a single $^{138}$Ba$^+$ atomic ion in a Paul trap, where $\Delta m$ is given by the difference in the magnetic quantum number of the atomic electron before and after the photon emission. Here, $\Delta m = 0$ corresponds to emission from a linear $\pi$ dipole and $\Delta m = \pm 1$ to emission from a circular $\sigma^{\mp}$ dipole.
Photons are emitted from the cooling transition, with $\lambda = 493.41\,$nm, (Extended Data Fig.~\ref{atom_exp}a). A magnetic field $B = 0.45\,$mT parallel to the axis of the trap ($z$ axis) defines the quantization axis perpendicular to the optical axis ($x$ axis). In the experimental sequence, we first Doppler cool the ion using a $493\,$nm cooling and a $650\,$nm repump laser in an axial trapping potential with frequency $\omega _z = 2\pi \times 660\,$kHz, reducing the extension of the motional atomic wavepacket down to $\sim 36\,$nm. This is followed by optically pumping to one of the Zeeman levels of the 6S$_{1/2}$ ground state. For example, when preparing a photon emission $\Delta m = +1$, we pump to the 6S$_{1/2},m_j = -1/2$ with a $\sigma^-$-polarized $493\,$nm laser and a repumper beam. Subsequently, we apply a short $\sigma^+$-polarized $493\,$nm laser pulse which excites the atom to the state 6P$_{1/2},m_j = +1/2$ (Extended Data Fig.~\ref{atom_exp}b,~d). From that excited state, the atom can spontaneously decay back to the 6S$_{1/2},m_j = -1/2$, through a $\Delta m = +1$ transition, to the 6S$_{1/2},m_j = +1/2$ through a $\Delta m = 0$ transition or to the 6D$_{3/2}$ manifold. During this transition the atom emits a photon that can be collected by the objective (NA = 0.40) and directed to the camera through the imaging system. To detect photons from the opposite transition ($\Delta m=-1$), the polarization of the optical pumping and excitation beams are exchanged as shown Extended Data Fig.~\ref{atom_exp}c. 

In this configuration, photons from $\Delta m = 0$ ($\Delta m=\pm1$) transition are horizontally (vertically) polarized along the optical axis. This allows us to select only photons from the $\sigma$ ($\Delta m=\pm1$) transitions by introducing a polarization beam splitter (PBS) after the objective so long as the aperture is small. An ideal PBS paired with the NA = 0.40 objective used here removes 99.998\% of photons from the $\pi$ transitions and 2.7\% of photons from the $\sigma$ transitions. The ratio between transmitted $\Delta m= 0$ and $\Delta m=\pm1$ photons is $\sim 10^{-4}$. We might therefore expect that the dipole image is not significantly changed by the polarization filtering at this numerical aperture, and indeed this is borne out by complete numeric simulations (see Extended Data Fig.~\ref{fig:plts1}). 

The light emitted during the cooling and optical pumping stages is filtered out by blocking the acquisition of the CCD sensor. The results shown in the main text were obtained using an intensified CCD camera (ICCD, Andor iStar A-DH334T-18H-63), with a pixel size of $13 \times 13\,\mu$m$^2$. This allowed us to gate the photo-intensifier when the preparation stage (cooling and pumping) was finished, enabling the collection of single photons only from the desired transition. Extended Data Fig.~\ref{atom_exp}d shows the sequence and timing used in the experiment.
 
\subsection*{Atom image characteristics, stability and drifts correction} 
The image of the atomic ion corresponds to the point-spread function of the imaging system which is well approximated in our case by a 2D Gaussian. After Doppler cooling, we image the single ion with the optical system described in the main text. The detected images are fitted to a Gaussian profile with seven free parameters $(z_0, y_0,\sigma_z,\sigma_y,A,O,\theta)$, being $(z_0,y_0)$ the coordinates of the centroids, $\sigma_z$ and $\sigma_y$ the standard deviation in the major and minor axis, $A$ the amplitude, $O$ an offset and $\theta$ rotation angle with respect to the CCD sensor axis. Although the Gaussian fitting method is in general informationally sub-optimal, it does not introduce significant errors in the measured NA$<n_{\text{med}}$ (where $n_{\text{med}}$ is the refraction index of that surround the emitter), and non-diffraction-limited regime \citeMethods{Stallinga2010M}.

The magnification of the imaging system is measured by imaging a string of two ions separated by well-known distance~\citeMethods{james1998quantum}, and is given by $M = 5.40(7)$.
The image of the atom detected on the CCD camera is a non diffraction-limited Gaussian profile, characterized by widths $2\sigma_z = 30.1(4)\,\mu$m and $2\sigma_y = 28.4(4)\,\mu$m, limited mostly by aberrations and misalignment of the in-vacuum optics. The expected displacement of $M\times 157.1\,$nm is 35 times smaller than the width of the images. 

The centroid of the images can be measured, in principle, with arbitrarily high precision as the number of collected photons increases~\citeMethods{bobroff1986,thompson2002M}. In our experiment, the overall photon detection rate is $\sim 1600$ photons/s and is limited by the sequence repetition frequency, numerical aperture of the system and quantum efficiency of the camera. Therefore, in order to achieve few nanometres precision it is necessary to accumulate photons for several hours. 

The long accumulation time introduces a new source of error in the position estimation that originates from mechanical drifts in the imaging system. The stability of the imaging system is characterized by the Allan variance of the fitted centroids of the detected images~\citeMethods{monroeadaptativeM}, which gives us a measure of the position uncertainty depending on the accumulation time $\tau$. This is done by taking $N$ pictures with exposure time t, adding them in bins of duration $\tau = n t$, where $n$ is a integer number smaller than $N/2$. Each binned image is fitted to the seven parameters Gaussian function, from where the centroids are extracted. For comparison, we also use, besides the ICCD camera, an EMCCD camera (Andor iXon DU-897) with bigger pixel size ($16\times 16\,\mu$m$^ 2$). In the case of the EMCCD camera we take 2000 images of $2\,$s exposure time with the atom emitting resonance florescence at maximum rate. In the case of the ICCD camera, we take 3000 images of $0.5\,$s exposure. In both cases, the time between two consecutive images is negligible. Extended Data Fig.~\ref{allan}a,~b shows the vertical position uncertainty extracted with this method. The minimum uncertainty in the vertical position obtained using the EMCCD camera is $2.13(41)\,$nm for $148\,$s accumulation time, while for the ICCD set-up the minimum is $3.29(71)\,$nm for $74\,$s accumulation time. In both cases the decreasing part of the curve is dominated by shot noise. The drift of the centre of the fitted reference images used in the experiment is shown in Extended Data Fig.~\ref{allan}c where we observe that in a period of $3\,$h the image drifts a maximum of $\sim 200\,$nm in both vertical and horizontal direction. To compensate for these drifts, we use the acquisition of long-exposure images during the cooling stage (Extended Data Fig.~\ref{atom_exp}e) to obtain a real-time `reference' of the particle position.
Extended Data Figs.~\ref{atom_exp}d,~e show the full experimental sequence. This sequence is repeated for $3\,$h, and the analysed pictures correspond to accumulation of photons in a $11 \times 11$ pixel sub-area of the CCD sensor. 

After the data collection is finished, each reference image is fitted, and the mean centroid position of two consecutive reference images is used to correct for the drifts in the signal image acquired between them. Then, we add up all the corrected signal images and we fit this data using the seven free parameters function. Finally, we compare the centroid positions of the added-up reference and signal images to determine their relative displacement. The uncertainty of the displacement is extracted from the $1\sigma$ confidence intervals using $\chi^2$ analysis, given its relation with the real noise sources~\citeMethods{bobroff1986}. The obtained value for the displacement in the object plane and its uncertainty are shown in Extended Data Fig.~\ref{error_evolution} as a function of the number of accumulated single images. 

A simplified analysis is done by considering the displacement between consecutive signal images, which correspond to the accumulations of $\Delta m = +1 $ or the $\Delta m = -1$ alternately. This analysis is valid since the time separation between this images is much shorter than the characteristic time of the mechanical drifts. These results are shown in Extended Data Fig.~\ref{histogram} and agree with the more precise analysis presented in the main text.

\section{Experimental methods in the nanoparticle experiment}

\subsection*{Set-up and sample preparation}
We deposit a single gold nanoparticle (BBI solutions, diameter $100\,$nm) on a silica nanofibre (diameter $410\,$nm) by touching the nanofibre with a droplet, provided by a syringe needle that contains a diluted suspension of nanoparticles. The deposition process is monitored using the same microscopy set-up also used for imaging the nanoparticle (without the solid immersion lens). The presence of a single nanoparticle on the nanofibre can be detected via absorption spectroscopy. The overall absorption allows us to detect if we deposited a single nanoparticle~\citeMethods{petersen2014chiralM}. In combination with the microscopy set-up it is possible to deposit and identify a single gold nanoparticle with a success probability close to one.

In order to place the nanoparticle in the centre of the solid immersion lenses, which are mounted on motorized three axis stages and can also be tilted, the lenses are aligned with respect to each other using the imaging system. Next, the two lenses are moved upwards such that the horizontal nanofibre fits into the gap between the lenses. Illuminating the nanoparticle via the nanofibre we then centre the lenses on the nanoparticle. Finally, the gap between the lenses is reduced to about $200\,\mu$m and filled up with immersion oil, which has the same refractive index as the nanofibre and the fused silica solid immersion lenses. In all the presented experiments the same nanoparticle was used.

The imaging system is a combination of a long working distance microscope and the solid immersion lens. The microscope consists of an infinity corrected objective by Mitutoyo, with a magnification of $20$ and an infinity tube lens to image onto a CCD camera (Matrix Vision mvBlueFOX3-1013G-2212). In two different measurements we used two different objectives with the numerical apertures of $\text{NA} = 0.28$ and $\text{NA} = 0.42$. The solid immersion lens is a half ball lens with a radius of $2.5$ mm. Via a surface topography standard, which provides periodic structures with precise known dimensions, we measure the magnification of the long working distance microscope. The overall magnification is then determined from the magnification of the immersion lens and the objective. This combination results in an overall imaging system with numerical apertures $\text{NA} = 0.41$ and $\text{NA} = 0.61$ and the magnifications $M_{0.41}=21.9(2)$ and $M_{0.61}=20.1(1)$.

\subsection*{Polarization adjustment}
In the experiment we use two laser beams: a reference and a measurement beam, with fixed and adjustable polarization respectively, see Fig.~\ref{setup}. The polarization of the reference beam is aligned along the $z$ axis using a Berek compensator. The measurement beam is set to be linear polarized along the $y$ axis before passing through a half- and then a quarter-wave plate. We align the optical axis of the two wave plates on the polarization of the measurement beam by adjusting each wave plate such that the incident polarization is not changed. 

By rotating the half-wave plate, which is mounted in a motorized rotation stage, we can adjust the beam's polarization to every elliptical polarization with the major axes along $x$ or $y$. This includes, \emph{e.g.}, the cases linear polarization parallel to the $x$ axis (imaging axis) (rotation angle: $\theta_\text{HWP}=0^{\circ}$), circular polarization ($\theta_\text{HWP}=+22.5^{\circ}$), linear polarization orthogonal to the imaging axis ($\theta_\text{HWP}=+45^{\circ}$) and opposite rotating circular polarization ($\theta_\text{HWP}=+67.5^{\circ}$). In the measurement sequences, we start with linear polarization orthogonal to the imaging axis and then we rotate the half-wave plate in positive direction by $90^{\circ}$. In order to avoid aberration caused by light propagating along the ridge of the two immersion lenses, the measurement beam is tilted by $7^\circ$ degrees from the $z$ axis, see Fig.~\ref{setup}b. This tilt is included in the theory plots shown in Fig.~\ref{results_image}f and Extended Data Fig.~\ref{Meth_Fig_02}b.

\subsection*{Data acquisition and analysis}
The illumination times of the images are $2\,$ms ($\text{NA}=0.41$ objective) and $6.5\,$ms ($\text{NA}=0.61$ objective, lower laser power). To reduce the effects of mechanical drifts, the pictures are taken alternately using the reference and measurement beams (Fig.~\ref{setup}), blocking either beam with a mechanical shutter. For every tuple we then determine the real (reference beam) and apparent (measurement beam) position of the nanoparticle. The apparent displacement of the particle is the difference between these measured positions.

In the experimental sequence, the particle displacements are measured as a function of polarization and the focal position of the imaging optics. For every polarization ratio $\epsilon$, the relative focus position is scanned by moving the long working distance microscope via a piezoelectric transducer, with a step size of $1.25\,\mu$m and a total range $\sim 20\,\mu$m. Then, the polarization ratio is changed by rotating the half-wave plate by $2.5^{\circ}$. 25 tuples of data are acquired for every $\epsilon$ and focus position.
Fig.~\ref{results_image} (main text) shows the mean displacements obtained from averaging over all displacements for the five focal positions closest to the focus of the imaging system. The statistical error of each data point displayed in Fig.~\ref{results_image}f is estimated as $\sigma_{\Delta y}/\sqrt{125}$, where $\sigma_{\Delta y}$ is the standard deviation of the measured displacements.

\subsection*{Position determination}
To correct for inhomogeneous pixel efficiencies of the CCD camera, we apply standard flat-field correction on the measured image data. Then, in order to determine the (apparent) position of the nanoparticle, we fit a 2D Gaussian with six free fit parameters to the particle images. The free parameters are the centroid position $(z_{0},y_{0})$, the amplitude $A$, the waists $\sigma_{z}$ and $\sigma_{y}$ of the elliptical Gaussian and an intensity offset $O$.

We check if the use of a 2D Gaussian function introduces a bias in the position determination by comparing different sets of fit parameters, \emph{i.e.}, with and without offset as well as fixed waists $\sigma_z$ and $\sigma_y$. We also fit the particle position using the experimentally obtained point-spread function of our imaging set-up~\citeMethods{Alberti2016M}. Within our experimental errors, all methods lead to the same displacements. In particular for $|\epsilon| \leq 2$, the difference in position obtained from the different methods is in the sub-nanometre regime. We therefore use 2D Gaussian fitting with six free parameters for all the data analysis shown in the main text.

\subsection*{Focus-dependent shifts}
In optical imaging, an additional shift of the apparent position of an emitter occurs when the light scattered by the particle is not homogeneously distributed along the aperture of the imaging system, and if the imaging system is not correctly focused, see Extended Data Fig. \ref{focus_dep_pos}. This effect (which has been studied previously \citeMethods{Lew2013,Enderlein2006M,Stallinga2010M,Engelhardt2011M,Backlund2012M,Toprak2006}) depends on the NA and, in contrast to the effect studied in this work, vanishes when the particle is perfectly in focus.

The difference in the origin of both polarization-dependent and focus-depends shifts can be understood from a wave-interference picture. A dipole rotating around the $z$ axis can be described as the superposition of $x$ and $y$ linear dipoles with a phase-difference of $\pi/2$. In this picture, the polarization-dependent shift of the centroid is a consequence of the interference of the doughnut-shaped field distribution of the longitudinal $x$ dipole (Eq.~(\ref{eq_image1})) and the Gaussian-type field distribution of the $y$ dipole (Eqs.~(\ref{eq_image2}) and (\ref{eq_image3})). In the case of perfectly focussed imaging, the two fields are in phase in the image plane and interfere maximally, leading to a maximum shift of the image centroid. When moving away from the focus, the two field distributions have different propagation phases, the interference contrast drops and the centroid shift reduces quadratically with the (small) distance to the focus.

On the other hand, if we consider a linear dipole emitter with a polarization axis which is not perpendicular or parallel to the optical axis, the field in the focal plane can again be decomposed into a Gaussian and a doughnut-shaped field distribution. These fields are $\pi/2$ out of phase in the plane of focus, and do not interfere. This, however, changes when the imaging system is sightly out of focus: due to the different phases acquired by propagation, the two fields partially interfere, and the shift of the centroid increases linearly with the defocusing.

Such an effect occurs in our experiment when the nanoparticle is illuminated via the measurement beam, which is tilted by $7^\circ$ from the $z$ axis. For $\epsilon\gg1$ the main dipole moment is tilted with respect to the optical axis, leading to an inhomogeneous distribution of the emitted light across the aperture of the imaging system. In our experiment, this focus-dependent shift occurs solely in $z$ direction. To compare this shift to the one discussed in the article, which solely occurs in $y$ direction, we measure the dependence of both shifts on the focal position of the objective. We scan the focal position of our imaging system and observe apparent shifts in the $z$ direction. This shifts increase approximately linearly with defocusing and are on the order of a few ten nanometres, depending on the polarization of the measurement beam, see Extended Data Fig.~\ref{Meth_Fig_02}a. On the contrary, we observe that, within the focal area, the angular momentum-dependent shift in the $y$ direction does not depend on the focal position within our measurement uncertainty, see Extended Data Fig.~\ref{Meth_Fig_02}b. This is expected from the predicted second-order dependence of the shift with the focal position. Moreover, this shift is an order of magnitude lager than the focus-dependent shift and therefore dominates the systematic localization error in our experiments.

\bibliographystyleMethods{naturemag}
\bibliographyMethods{library}
\clearpage

\renewcommand{\figurename}{Extended Data Figure}
\setcounter{figure}{0} 
\renewcommand\thesubsection{\arabic{subsection}}
\setcounter{page}{1}

\onecolumngrid

\begin{figure}[h]
\textbf{\large Extended data, figures and tables}
\vspace{5mm}

\centerline{\includegraphics[width=160mm]{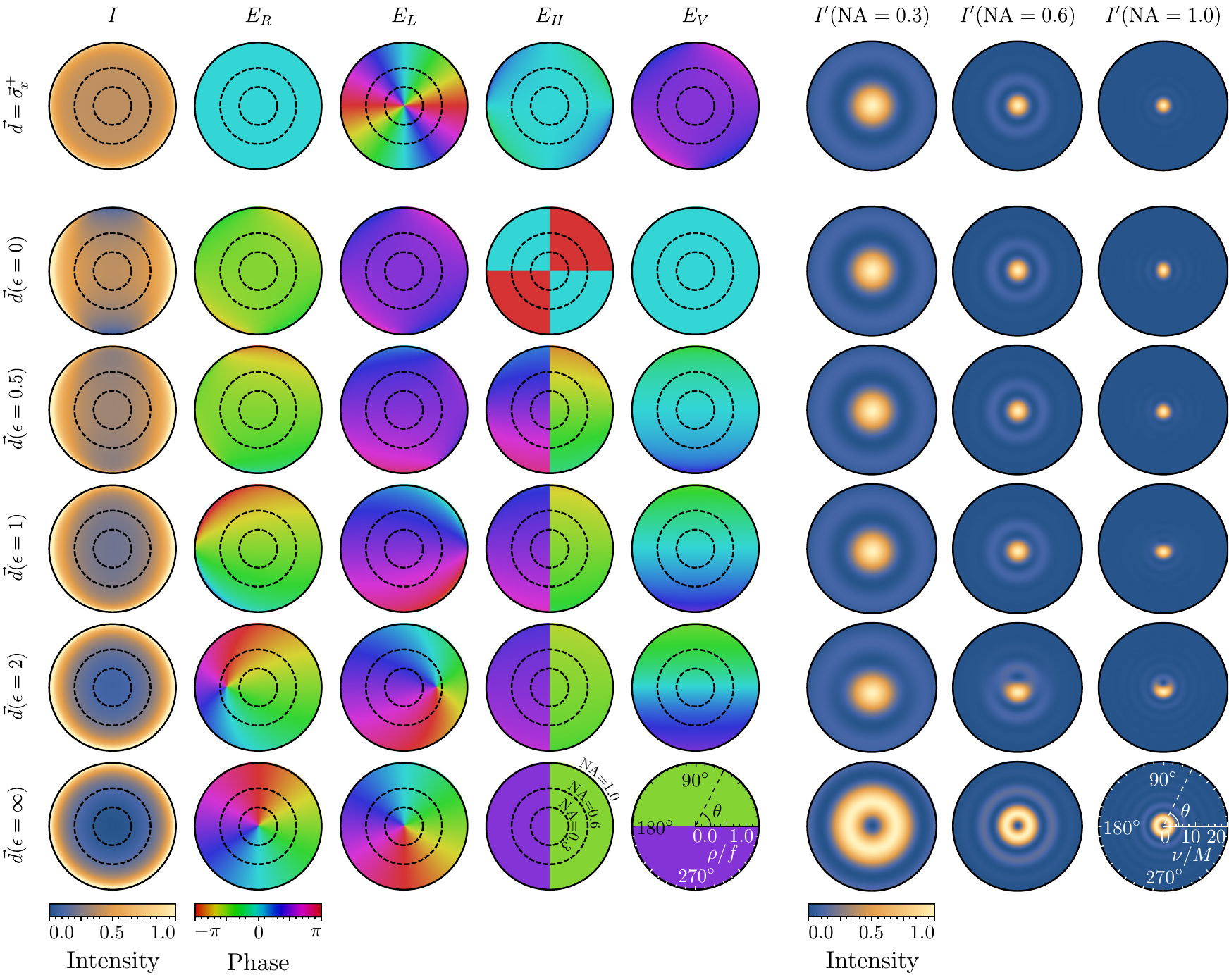}}
\caption{\textbf{Dipole fields and images: Momentum vortices and weak value amplification.} There is a close connection between the observed weak value amplification and the appearance of momentum vortices in the emitted light field.The plots show the field distribution of the emitted light at the lens plane for different polarization states of the emitter alongside corresponding intensity distributions in the image plane. 
\textbf{(Left five columns)} Field intensity $I$ and phase in the local linear ($E_H$,$E_V$) and local circular ($E_R$, $E_L$) polarization bases of the dipoles shown at left by an apodized orthographic projection. This projection is identical to the field distribution after collimation by an ideal spherical lens. The fields are plotted in radial coordinates $\rho/f = \sin{(\phi)}$ to an aperture half-angle $\phi_a = \pi/2$ at which the orthographic projection diverges. Dashed circles indicate NA$= 0.3$, $0.6$. \textbf{(Right three columns)} Corresponding images $I'$ calculated by full propagation of the optical dipole fields for NA$= 0.3$, $0.6$ and $1$. For NA$<1$ the images are calculated by truncated Hankel transform. $I'$ is plotted in radial coordinates $\nu/M$ with units $\lambda / 2 \pi$. The colour scale is normalized to the maximum of each image. \textbf{(Row 1)} the dipole circularly polarized about the optical axis. \textbf{(Rows 2--6)} dipoles with increasing polarization ratio $\epsilon$. The corresponding images for negative $\epsilon$ can be obtained by reflecting the images along the horizontal axis. Optical spin-orbit coupling manifests in the azimuthal phase of axially symmetric dipole fields. For example, the right circularly polarized dipole about the optical axis (row 1) is a superposition of a right circular polarized field with orbital angular momentum $\hat{L}_x = 0$ and a left circular polarized field with $\hat{L}_x = 2 \hbar$. Similarly, the linear dipole along the optical axis, $\vec{d}(\epsilon = \infty)$, consists of equal superposed circular fields with orbital angular momentum $\hat{L}_x = \pm \hbar$ opposed to their spin $\hat{S}_x = \mp \hbar$. For $0<\epsilon<\infty$ the $E_V$ component of $\ket{\psi}$ has a pronounced vertical phase gradient due to orbital angular momentum $\hat{L}_z$ that Fourier transforms to displacement along $y$ in the image plane. The image fields are both displaced and distorted depending on ellipticity and aperture. For small numerical apertures the image of the elliptical dipole is close to a displaced aperture point-spread function. 
For $1<\epsilon<\infty$ the circular components of the elliptical dipole fields have off-axis momentum-current vortices (phase singularities in the $E_R$ and $E_L$ plots) \cite{berry2011lateral,barnett2013,arlt2003handedness}. As shown in columns 2 and 3, the phase singularity moves from the edge ($\rho / f = 1$) to the optical axis ($\rho / f = 0$) as $\epsilon$ increases from 1 to $\infty$. The image plane distributions provide a graphical illustration of the apparent position shift. The elliptical dipoles may be displaced by an amount $\Delta y$ that corresponds to momentum larger than the field's momentum eigenmode spectrum and that scales inversely with the NA, according to the weak value amplification rule (see Extended Data. Fig.~\ref{displacements} and Eq. (4) in the main text). This weak value amplification is an example of supermomentum in single-photon field. The centroid is maximally displaced when the vortices are in the edge of the collection aperture, which in the case of NA $\sim$ 0.6 occurs when $\epsilon\sim 2$. }
\label{many_images}
\end{figure}

\begin{figure}
\centerline{\includegraphics[width=0.70\textwidth]{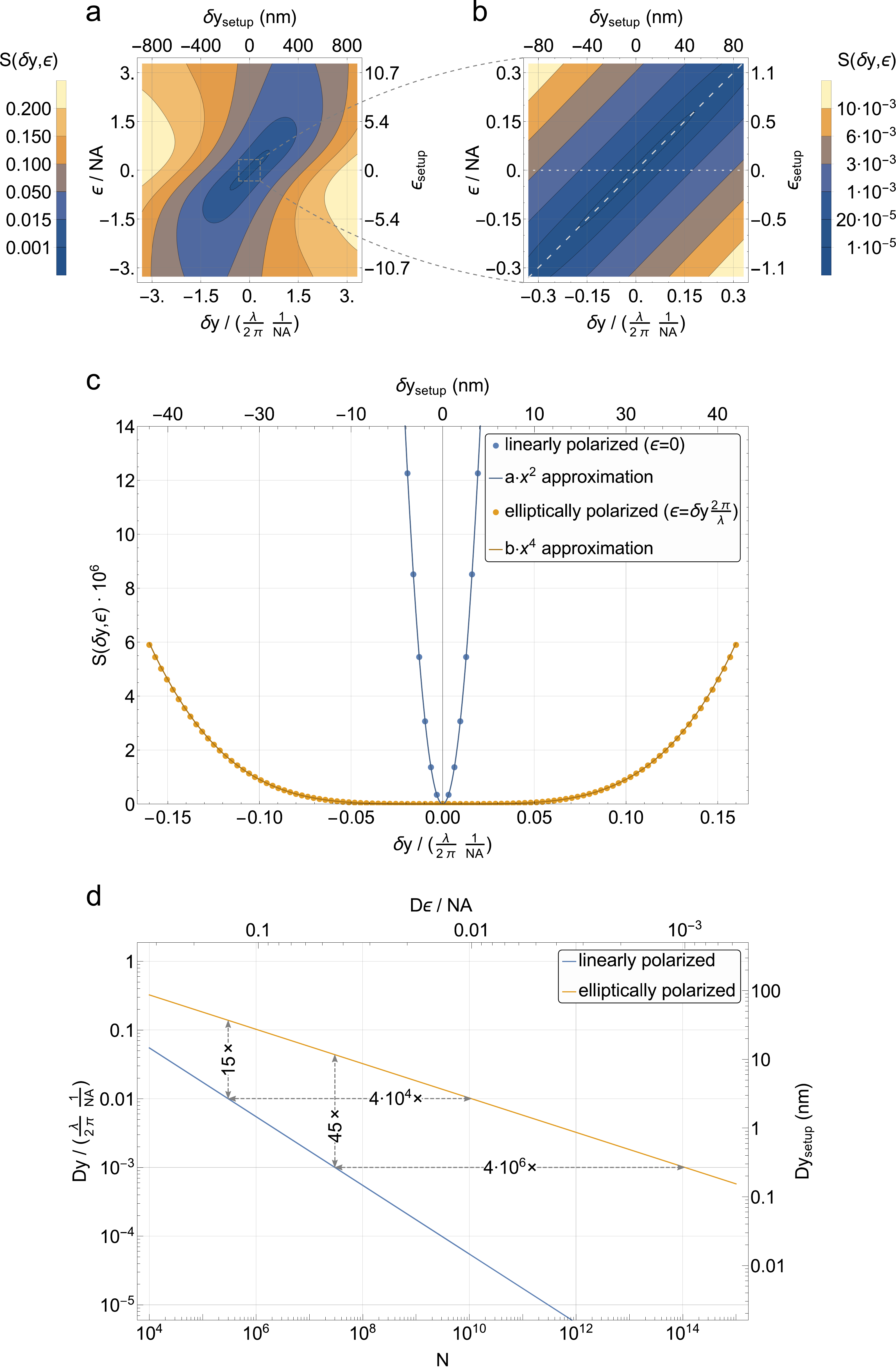}}
\caption{\textbf{Limits in the precision of position estimation}
\textbf{a}, Two dimensional plot of the function $S(\delta y, \epsilon)$ from Eq.~(\ref{eqn:1}), where the lower $x$-axis represents the real shift $\delta y$ of the emitter in the object plane in the normalized units $\lambda / (2\pi \cdot \text{NA})$ and the left $y$-axis represents the dipole polarization ratio $\epsilon$ in units of the $\text{NA}$. The upper $x$-axis and right $y$-axis represent $\delta y$ and $\epsilon$ corresponding to the set-up of the nanoparticle experiment with the overall $\text{NA} = 0.41$. \textbf{b}, A zoom of the centre region of \textbf{a}. \textbf{c}, $S$ evaluated along $\epsilon = 0$ (blue dots, dotted grey line in \textbf{b}) and along $\epsilon = \delta y \cdot 2\pi / \lambda$ (yellow dots, dashed grey line in \textbf{b}). For small $\delta y$ and $\epsilon$ the function $S$ can be approximated by $S(\delta y, \epsilon)\approx a\cdot \delta y^{2}$ for $\epsilon=0$ and $S(\delta y, \epsilon)\approx b\cdot \delta y^{4}$ for $\epsilon=\delta y \cdot 2\pi / \lambda$. The fitted values are $a=33.3\cdot 10^{-3}$ and $b=9.1\cdot 10^{-3}$. \textbf{d}, Precision limit $Dy$ for the particle localisation resulting from photon shot noise for solely linearly polarized emitters (blue curve) and slightly elliptically polarized emitters (yellow curve) as a function of the collected photon number $N$. For a solely linearly polarized emitter, the precision increases much faster with $N$ than for the case where elliptical polarization can be present. The latter case requires orders of magnitude larger photon numbers to gain the same precision, as indicated by the horizontal dashed lines. The vertical dashed lines indicate the precision limit increase for a two given photon numbers.
The upper $y$-axis shows the residual uncertainty $\delta \epsilon$ of the dipole polarization ratio of the imaged emitter.
}
\label{Precision_limit_Fig}
\end{figure}

\begin{figure}
\centerline{\includegraphics[width=0.95\textwidth]{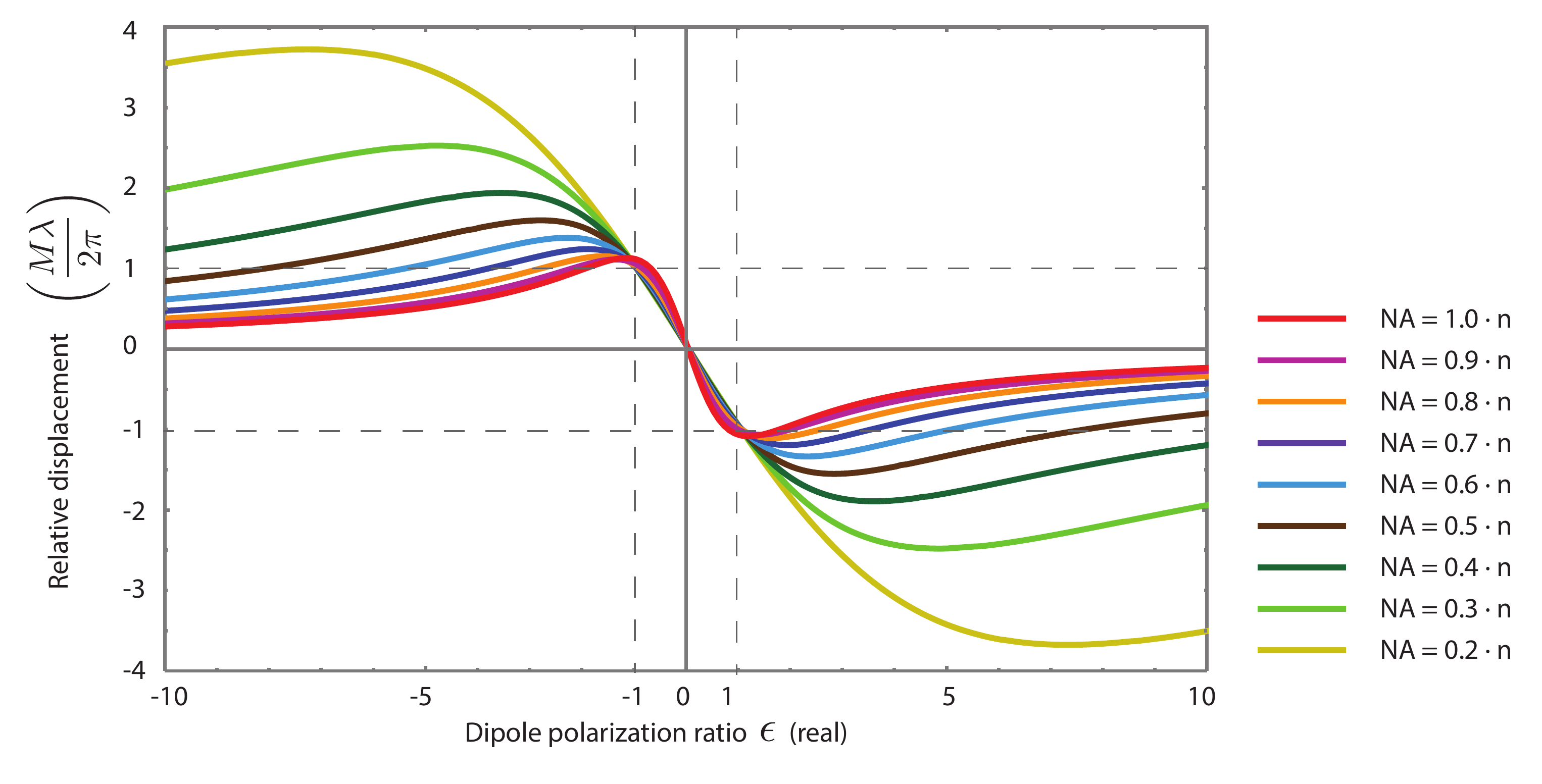}}
\caption{\textbf{Predicted centre of mass shift without small NA approximations.}
Predicted apparent displacement of the centre of mass of the images of dipoles for different NA as a function of the dipole polarization ratio $\epsilon$, for real $\epsilon$, \emph{i.e.}, elliptical rotating dipole calculated using full field propagation. The dashed lines show that for clockwise and anticlockwise circular rotating dipoles ($\epsilon = 1$ and $\epsilon= -1$ respectively), the displacement of the centre of mass of the image correspond to $M\lambda/(2\pi)$ and $-M\lambda/(2\pi)$ respectively.
}
\label{displacements}
\end{figure}

\begin{figure*}[h]
\centerline{\includegraphics[width=0.7\textwidth]{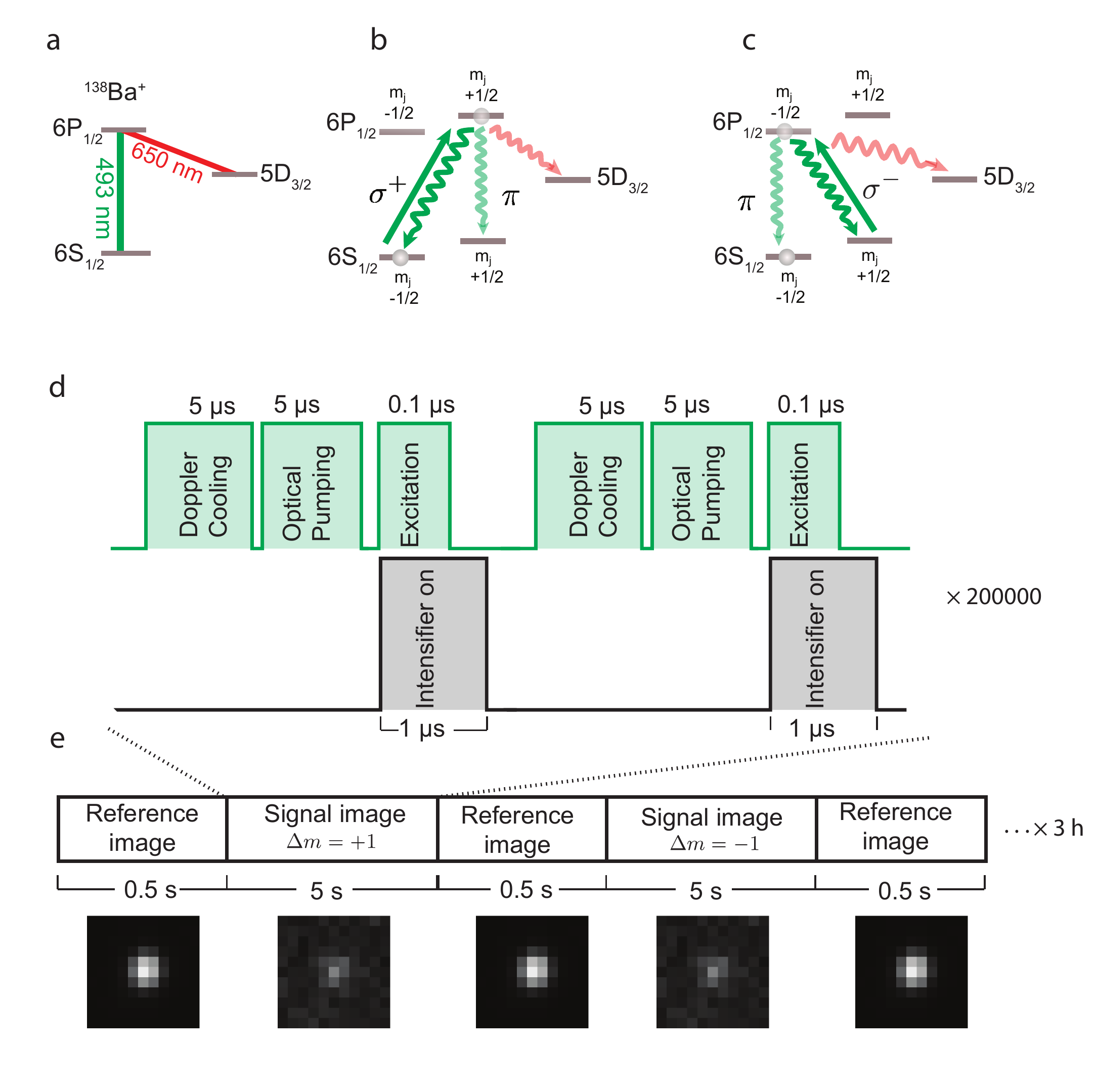}}
\caption{\textbf{Barium electronic states and experimental sequence.} 
\textbf{a,} Electronic states and transitions of $^{138}$Ba$^+$. \textbf{b,} In order to emit a $\Delta m=-1$ photon, the atom is prepared in the state 6S$_{1/2},m_j=-1/2$ and then excited by a circularly polarized ($\sigma^+$) laser beam. \textbf{c,} In order to emit a $\Delta m=+1$ photon the atom is prepared in the state 6S$_{1/2},m_j=+1/2$ and then excited by a circularly polarized ($\sigma^-$) laser beam. Other possible decays are filtered out. \textbf{d,} Timing of the sequence used for the generations of photons from a a given transition. The sequence is repeated 200000 times which results in a total duration of $5\,$s. \textbf{e,} A reference image of $0.5\,$s is taken before and after the 5 s accumulation of the desired photons. The acquisition is directly alternated between images with photons coming from the $\Delta m = -1$ and $\Delta m = +1$ transitions. 
}
\label{atom_exp}
\end{figure*}

\begin{figure}[h]
	\centering
	\includegraphics[width=0.8\textwidth]{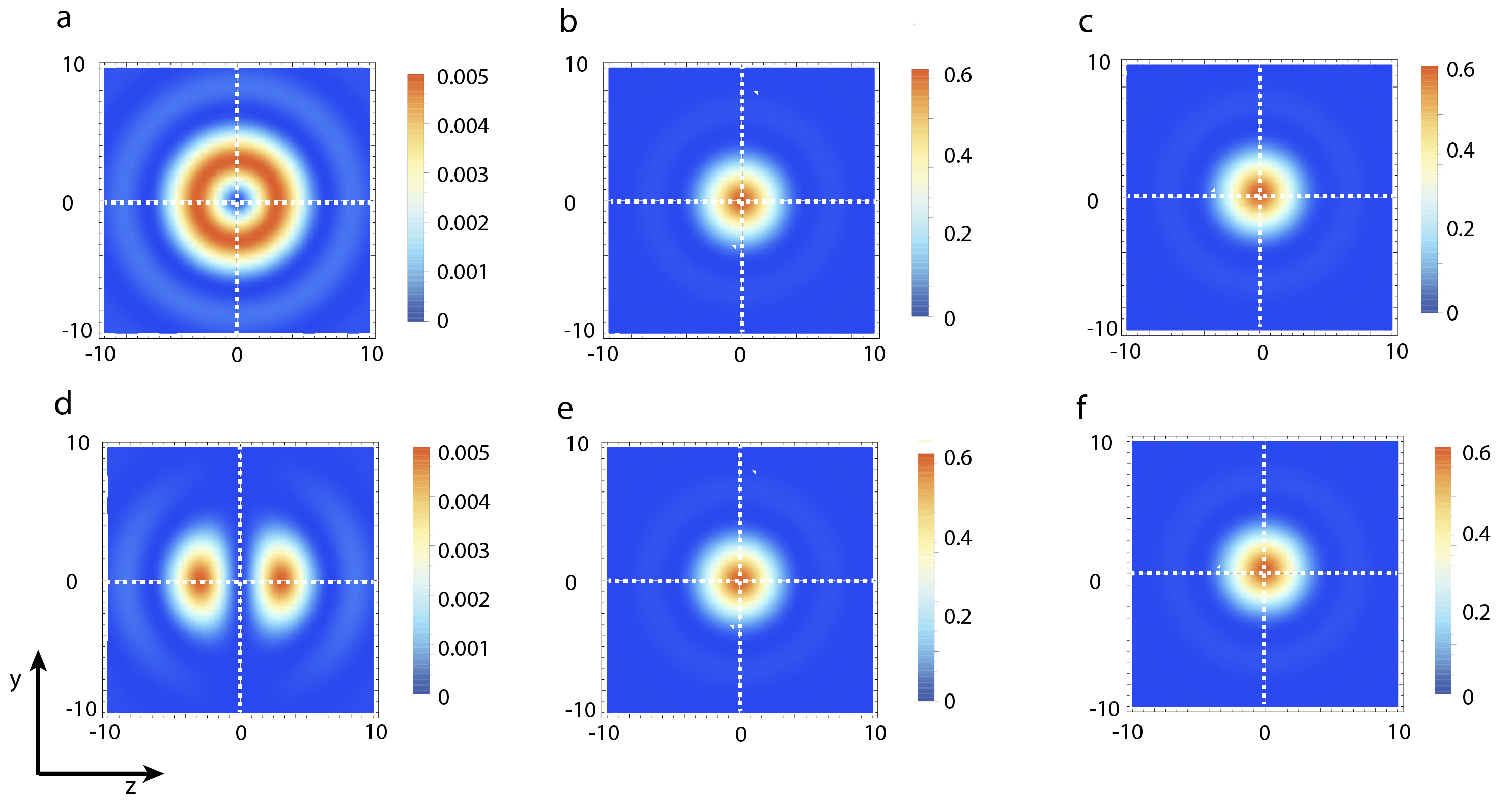}
	\caption{\textbf{Effect of polarization filtering in the position of the centroid of the image.}
Calculated intensity distribution in the image plane for a dipole oriented along the optical axis \textbf{a}, oriented orthogonal to the optical axis \textbf{b}, and a circular dipole which is the superposition of them, \textbf{c}. \textbf{d, e} and \textbf{f} show the image of these dipoles when placing a PBS between the two lenses forming the imaging system. The contribution of the longitudinal dipole (\textbf{a} and \textbf{d}), as can be seen from the colour scale, is orders of magnitudes smaller than the contribution of the transversal one (\textbf{b} and \textbf{e}). While introducing a PBS changes the intensity distribution of the image of the dipole polarized along the optical axis, it does not change the overall shape and displacement of the total image. Vertical and horizontal axes are in units of $M\lambda/2\pi$.
}
\label{fig:plts1}
\end{figure}

\begin{figure}
\centerline{\includegraphics[width=0.95\textwidth]{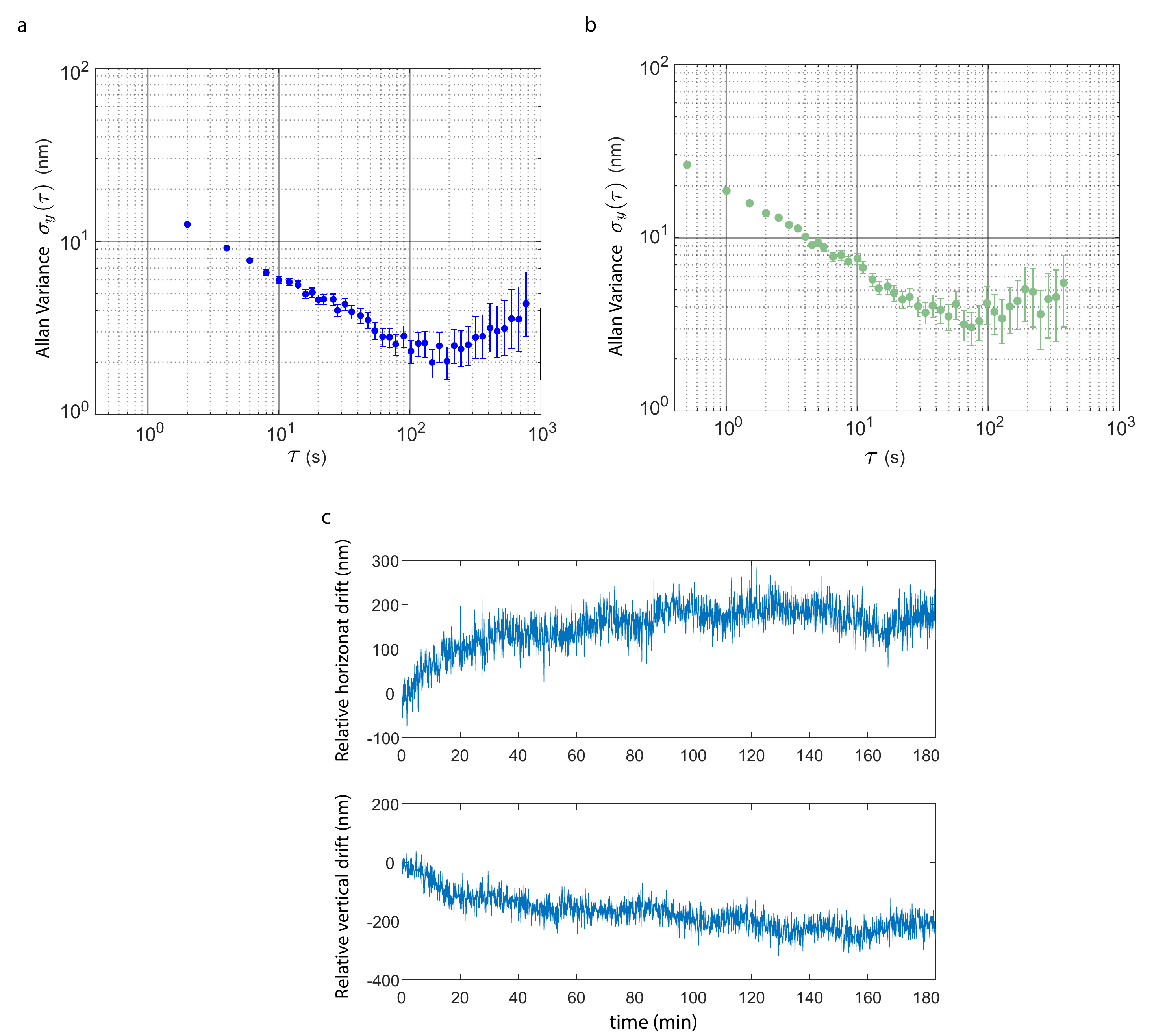}}
\caption{\textbf{Stability of the atom imaging system.} 
Allan variance of the vertical position the atom as a function of the accumulation time $\tau$ of the images, when using a EMCCD camera (\textbf{a}) and a ICCD camera (\textbf{b}). \textbf{c} Vertical and horizontal positions of the centroid of the fitted reference images, over a period of $180\,$min, using the ICCD camera.
}
\label{allan}
\end{figure} 

\begin{figure*}
\centerline{\includegraphics[width=0.6\textwidth]{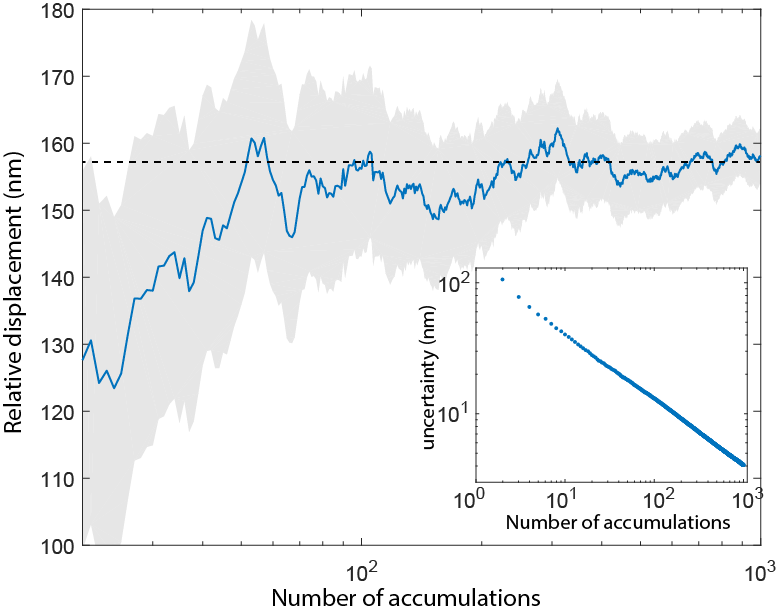}}
\caption{\textbf{Convergence of the measured relative displacement of the atom.} Convergence of the estimated relative displacement between the counter rotating circular atomic dipoles $\sigma^+$ and $\sigma^-$ (blue line) and uncertainty (grey area) versus the number of signal images accumulated after drifts correction. The dashed line shows the expected value $\lambda_1/\pi = \text{157.1 nm}$. The inset shows the evolution of the uncertainty in logarithmic scale as the number of accumulated images increases. 
}
\label{error_evolution}
\end{figure*}

\begin{figure}
\centerline{\includegraphics[width=0.6\textwidth]{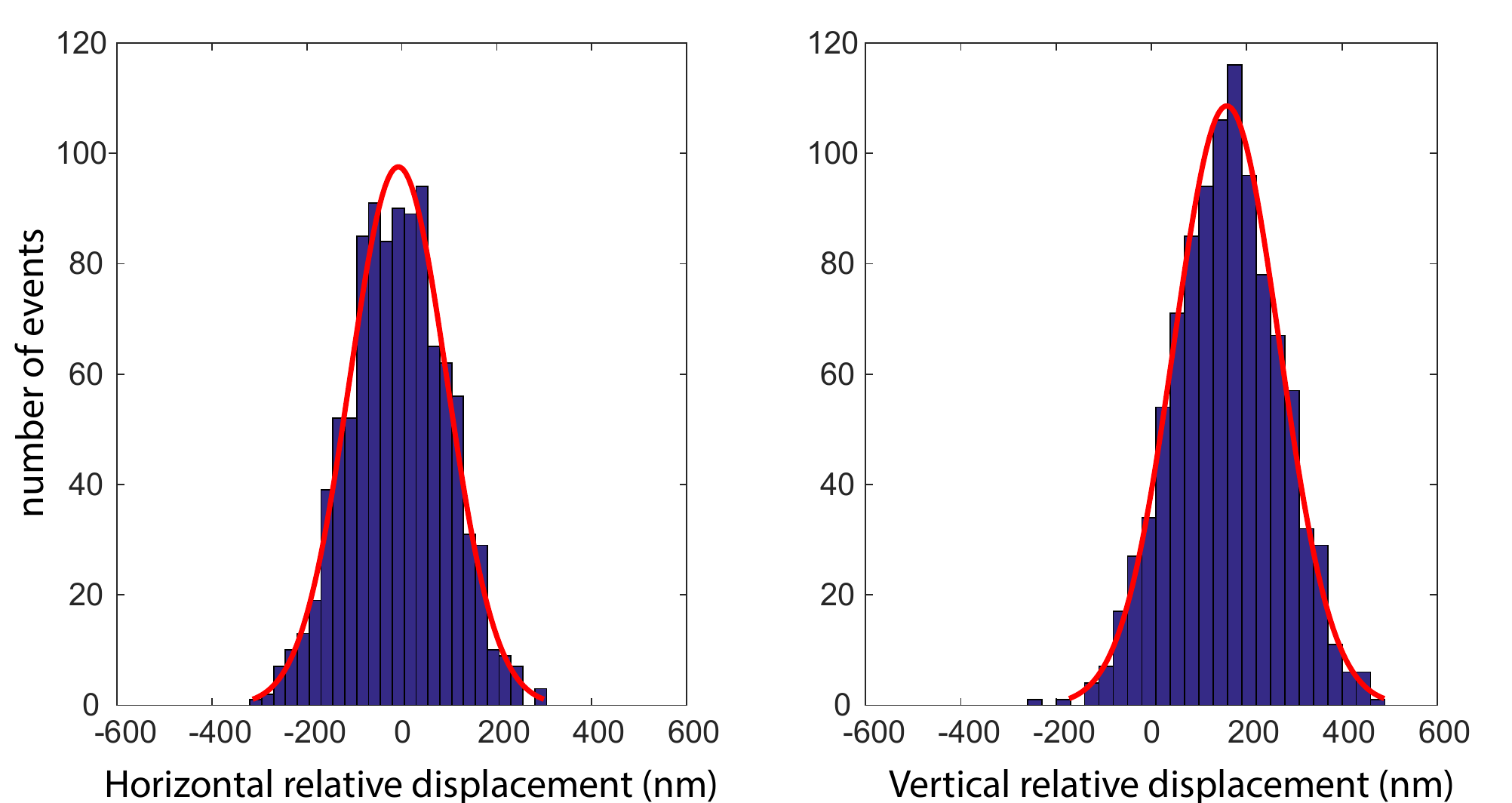}}
\caption{\textbf{Direct comparison of consecutive atom images}
Histograms of the horizontal and vertical relative displacements of pairs of consecutive signal images formed by photons coming from the $\Delta m = -1$ and $\Delta m = +1$ atomic transitions, for 2000 image pairs. The histograms show a clear average vertical displacement, and an average zero horizontal displacement. Each histogram is fitted to a normal distribution from which we extract an average horizontal relative displacement of $7(6)\,$nm and standard deviation of $106(4)\,$nm, while for the vertical displacement we get an average of $158(6)\,$nm and standard deviation of $112(4)\,$nm. The stated errors correspond to the $1\sigma$-confidence intervals of the fits.}
\label{histogram}
\end{figure}

\begin{figure}
\centerline{\includegraphics[width=0.8\textwidth]{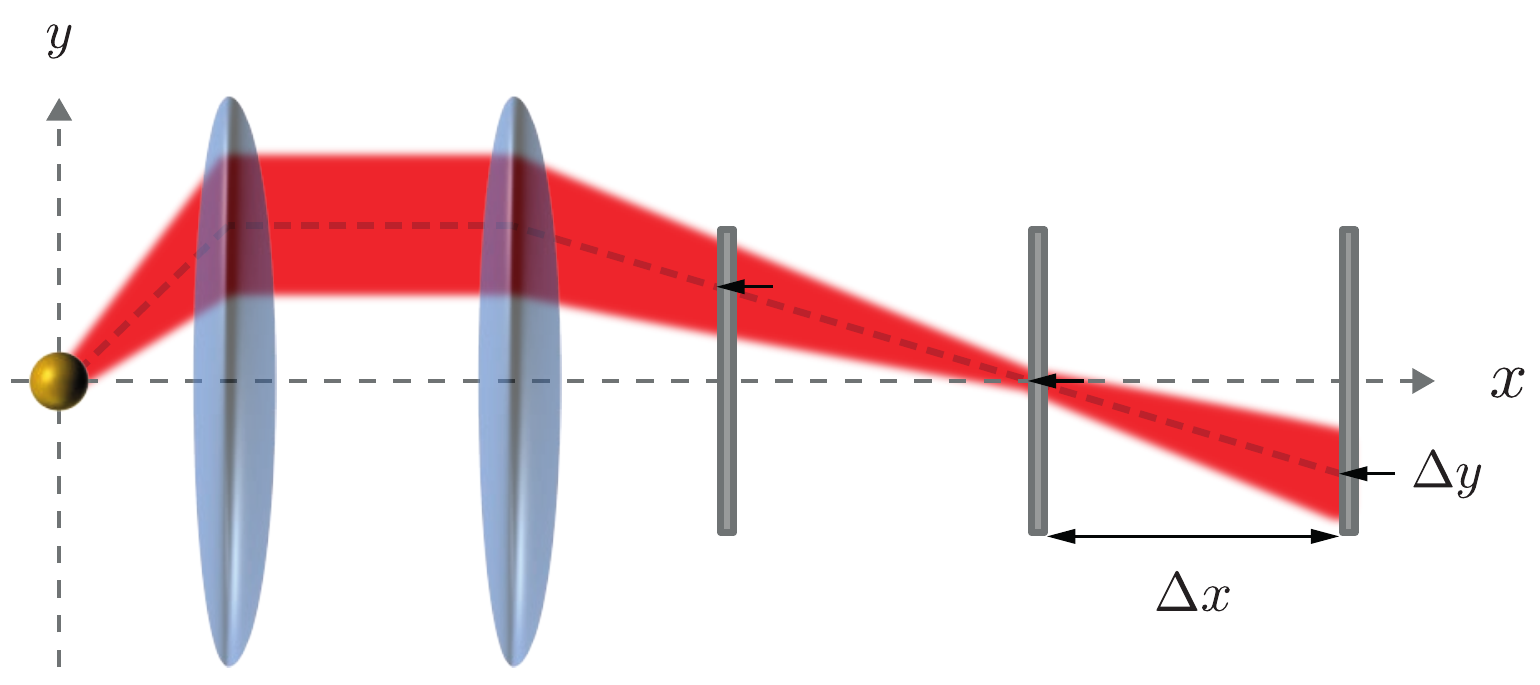}}
\caption{\textbf{Basic scheme of the focus-dependent displacement.} The red dotted line points out the vertical centre of intensity of the propagating beam, which inhomogeneously illuminates the imaging optics. When perfectly in focus (centre screen) no additional shift occurs. But on a screen at a position $\Delta x$ with respect to the focus, a shift of $\Delta y$ arises which depends linearly on $\Delta x$.}
\label{focus_dep_pos}
\end{figure}

\begin{figure}
\centerline{\includegraphics[width=0.8\textwidth]{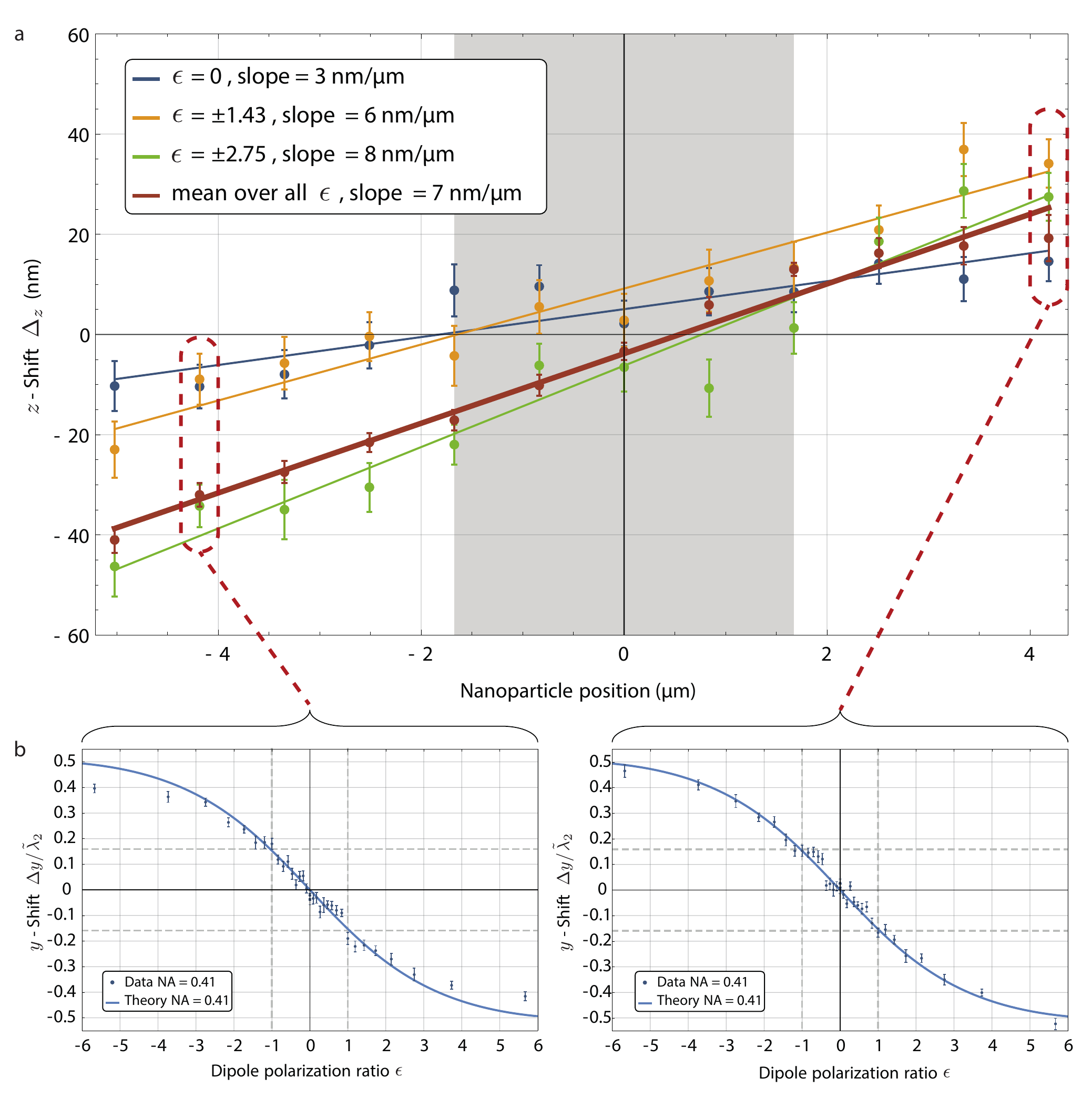}}
\caption{\textbf{Focus-dependent displacement.}
\textbf{a}, Experimental data showing the apparent displacement of the nanoparticle in the $z$ direction as a function of its distance to the focal point of the imaging system. The blue, yellow and green data correspond to the three polarization ratios $\epsilon = 0, \pm 1.43$ and $\pm 2.75$. Polarization ratios with opposite signs lead to the same inhomogeneous illumination across the aperture of the imaging system and thus give rise to the same shift. The solid lines are linear fits with nanoparticle displacement slopes of $3\,$nm, $6\,$nm and $8\,$nm per $\mu$m of defocusing, respectively. The red dataset is the averaged $z$ displacement over all polarization ratios $\epsilon$ with a slope of $7\,$nm per $\mu$m of defocusing. Error bars correspond to the statistical error. The error bars of the blue, yellow and green data sets are larger than for the red data set since they are averaged over 50 measurements per point, whereas the red dataset is averaged over 850 measurements. For $\epsilon=0$, we still observe an increase in displacement with a non-zero slope. This could be explained by a residual inhomogeneity in the illumination of the objective which could originate from a small tilt of the reference beam on the order of $0.1^{\circ}$. 
The data presented in the main text was taken from the measurements that lie within the grey region of the graph, which indicates the focal region of our imaging system. \textbf{b}, $y$ displacement as a function of the dipole polarization ratio $\epsilon$ plotted for the two cases where the nanoparticle is $\pm 4.2\,\mu$m out of focus. The data fits well to the theory curve and does not show any additional displacement in the $y$ direction, which would be visible as vertical offset.}
\label{Meth_Fig_02}
\end{figure}

\clearpage

\renewcommand{\figurename}{Methods Figure}
\setcounter{figure}{0} 
\setcounter{equation}{0}
\setcounter{page}{1}
\renewcommand\thesubsection{\arabic{subsection}}

\thispagestyle{empty}
\onecolumngrid
\begin{center}
\vspace{5 mm}
\textbf{\large Supplementary Information for:\\Wavelength-scale errors in optical localization due to spin-orbit coupling of light}\\
\vspace{2 mm}
G. Araneda,$^1$ S. Walser,$^{2}$ Y. Colombe,$^1$ D. B. Higginbottom,$^{1,3}$ J. Volz,$^{2}$	R. Blatt,$^{1,4}$ and A. Rauschenbeutel$^{2}$\\
\vspace{2 mm}

\textit{\small
$^1$Institut f\"{u}r Experimentalphysik, Universit\"{a}t Innsbruck, Technikerstra\ss e 25, 6020 Innsbruck, Austria\\
$^2$Vienna Center for Quantum Science and Technology, TU Wien-Atominstitut, Stadionallee 2, 1020 Vienna, Austria\\
$^3$Centre for Quantum Computation and Communication Technology, Research School of Physics and Engineering, The Australian National University, Canberra ACT 2601, Australia\\
$^4$Institut f\"{u}r Quantenoptik und Quanteninformation, \"{O}sterreichische Akademie der Wissenschaften, Technikerstra\ss e 21a, 6020 Innsbruck, Austria\\
}

\vspace{10 mm}
\end{center}
\twocolumngrid
\renewcommand{\figurename}{Supplementary Figure}
\setcounter{figure}{0}
\setcounter{section}{0}
\renewcommand\thesubsection{\arabic{subsection}}

\section{Wavefronts of the radiated field}
In the far field $|\boldsymbol{r}|\gg\lambda$, the electric field emitted by an optical dipole oscillating with frequency $\omega$ is
\begin{equation}
\boldsymbol {E}(\boldsymbol r,t) = -\frac{\omega^2}{4\pi\epsilon_0c^2}
\frac{e^{i( k r-\omega t)}}{r^3} (\boldsymbol{r}\times\boldsymbol{\mu})\times \boldsymbol{r}
\end{equation}
where $\boldsymbol{\mu}=\mu\boldsymbol{e}_\mu$ is the complex vector amplitude of the electrical dipole and $\mu=|\boldsymbol{\mu}|$. The quantity $r=|\boldsymbol{r}|$ is the distance to the dipole located at $r=0$ and $k=2\pi/\lambda$ is the wavevector of the emitted light with wavelength $\lambda$. Any dipole radiation can be decomposed into the radiation of the linear dipole $\boldsymbol{e}_0=\boldsymbol{e}_z$ oscillating along $z$ and the radiation of the circularly polarized dipoles $\boldsymbol{e}_\pm=\mp 1/\sqrt{2}(\boldsymbol{e}_x\pm i\boldsymbol{e}_y)$ that rotate in the $x-y$ plane. Here, $\boldsymbol{e}_x$, $\boldsymbol{e}_y$ and $\boldsymbol{e}_z$ are the unit vectors along the $x$, $y$ and $z$-axis, respectively. The corresponding fields in spherical coordinates are then
\begin{eqnarray}
\boldsymbol{E}_0(\boldsymbol{r})&=&\frac{\mu\omega^2}{4\pi\epsilon_0c^2}\sin\theta e^{i(k r -\omega t)}\boldsymbol{e}_\theta\\
\boldsymbol{E}_\pm(\boldsymbol{r})&=& \frac{\mu\omega^2}{4\sqrt{2}\pi\epsilon_0c^2}\frac{e^{i(k r -\omega t\pm \phi)}}{r}\left(
e^{i\frac{\pi}{2}}\boldsymbol{e}_\phi\mp\cos{\theta}\boldsymbol{e}_\theta\right)
\end{eqnarray}
where $r$, $\theta$ and $\phi$ are spherical coordinates with the corresponding unit vectors $\boldsymbol{e}_r$, $\boldsymbol{e}_\theta$, $\boldsymbol{e}_\phi$, respectively. We now analyse the field emitted in the $x-y$ plane $(\theta=0)$. The wavefronts, \emph{i.e.}, the points where the field has a given phase, are described by the parametric equation
\begin{eqnarray}
	r_0(\phi)&=&\frac{\omega t}{k}+\text{const}. \label{eqn_lin}\\ 
	r_\pm(\phi)&=&\frac{\mp\phi+\omega t}{k}+\text{const}. \label{eqn_spiral}
\end{eqnarray}
The phase fronts generated by a linear dipole, described in Eq. \ref{eqn_lin}, correspond to those of a spherical wave, similar to the case of a scalar (longitudinal) wave. On the other hand, the wavefronts emitted by a circular dipole are Archimedean spirals rotating around the $z$ axis, with a direction of rotation that depends on that of the dipole.

\section{Apparent position of the dipole}

\subsection{Geometric interpretation of the displacement}

The fields emitted by the $\boldsymbol{e}_+$ and $\boldsymbol{e}_-$ dipoles have spiral wavefronts in the $x-y$ plane, with opposite orientations. At a given location in that plane, the direction of propagation $\boldsymbol{k}$ of the photons is given by the normal to the wavefronts, which is slightly tilted with respect to the radial direction. The tilt angle $\gamma(r)$ can be obtained from the geometric scheme shown in Extended Figure 3. For a small increase in the angle $\delta\phi$, the distance of the wavefront to the origin increases by $\delta r = \mp\delta \phi /k$ (see Eq. \ref{eqn_spiral}). Consequently, we obtain for the tilt angle 
 $\gamma = \delta r/ r\delta\phi = \mp1/kr$. When the particle is observed along the $x-y$ plane, this angular tilt translates into a transverse shift of $\delta s = \mp1/k$ of the apparent origin of the wavefronts from the actual position of the dipole. 

\subsection{Calculating the image}
In order to get a more rigorous prediction of the apparent position of the particle we consider a typical microscopy set-up, where we image the particle using, \emph{e.g.}, two lenses with focal length $f$ (object side) and $f'$ (image side). For the object, we assume it to be located in a medium with refractive index $n$ whereas on the image side we assume a medium with refractive index $n'$. Furthermore, we set the optical axis of our microscope to be aligned with the $z$-axis of our coordinate system (note that, for simplicity, we used a different optical axis than the one in the main text). In order to obtain the apparent particle position, we calculate the point spread function, \emph{i.e}, the intensity distribution of the light emitted by the dipole in the image plane of our microscope. Up to a global phase factor, this field distribution of a dipole in the image plane of our microscope is given by \citeSupplement{novotny2012principles} 
\begin{eqnarray}
\boldsymbol{E}(\rho,\varphi)=\frac{\omega^2}{8\pi\epsilon_0^2c^2}k'\boldsymbol{G}\cdot\boldsymbol{\mu}
\end{eqnarray}
where $\rho$ and $\varphi$ are the polar coordinates in the image plane, $k'=2\pi/\lambda n'$ is the wavevector on the image side and the Green's function $\boldsymbol{G}$ is given by 

\begin{align}
\boldsymbol{G}=&\frac{f}{f'}\sqrt{\frac{n}{n'}}\cdot \nonumber\\
&\left(
\begin{array}{ccc}
	I_{0}+I_{1}\cos2\varphi & I_2 \sin2\varphi & -2iI_1\cos \varphi \\
	I_{2}\sin2\varphi & 	I_{0}-I_{1}\cos2\varphi & -2iI_1\sin \varphi \\
0 & 0 & 0 
\end{array}	
\right)\label{eqn_image}
\end{align}
The integrals are given by
\begin{eqnarray}
I_0&=&\int_{0}^{\theta_{m}}d\theta\sqrt{\cos\theta}\sin\theta(1+\cos\theta)J_0(x) \label{eqn_int0} \\
I_1&=&\int_{0}^{\theta_{m}}d\theta\sqrt{\cos\theta}\sin^2\theta J_1(x) \label{eqn_int1} \\
I_2&=&\int_{0}^{\theta_{m}}d\theta\sqrt{\cos\theta}\sin(1-\cos\theta)J_2(x)\label{eqn_int2}
\end{eqnarray}
where $J_i(x)$ are the Bessel functions of order $i$ with the argument $x=k'\rho\sin\theta f/f'$ and $\theta_{m}$ is the maximum opening angle of the imaging system and depends on the numerical aperture by $\text{NA}=n\sin\theta_{m}$.

In order to derive an analytic expression for the image of the particle, we limit the following calculations to the case of small aperture in the image size, \emph{i.e.}, use the approximations $\cos\theta\approx1$ and $\sin\theta\approx\theta$. Consequently, the integrals in Eqs. \ref{eqn_int0}-\ref{eqn_int2} simplify to
\begin{eqnarray}
I_0&=&\int_{0}^{\theta_{m}}d\theta 2\theta J_0(x)=\frac{2\theta_{m}}{k'\rho f/f'}J_1(\tilde{\rho})\\ \label{eqn_int0b}
I_1&=&\int_{0}^{\theta_{m}}d\theta\theta^2 J_1(x)=\frac{\theta_{m}^2}{k'\rho f/f'}J_2(\tilde{\rho})\\ \label{eqn_int1b}
I_2&=&0\label{eqn_int2b}
\end{eqnarray}
with $\tilde{\rho}=\rho\cdot k'\theta_{m}f/f'$ and we obtain for the Green's function:
\begin{align}
	\boldsymbol{G}&=&2\sqrt{\frac{n}{n'}}\frac{\theta_{m}}{k'\rho} \cdot
	\left(
	\begin{array}{ccc}
		J_1 & 0 & -i\theta_{m}J_2\cos\varphi \\
		0 & J_1 & -i\theta_{m} J_2\sin\varphi \\
		0 & 0 & 0 
	\end{array}.	
	\right)\label{eqn_image2}
\end{align}

The final field distribution in the image is a superposition of the distributions of the three dipoles oscillating along $x$, $y$ and $z$ direction. The electric field distributions of these dipoles in the image plane are given by
\begin{eqnarray}
\boldsymbol{E}_x&=&E_0\cdot \frac{\theta_{m}}{\rho} J_1(\tilde{\rho})\boldsymbol{e}_x\\
\boldsymbol{E}_y&=&E_0\cdot \frac{\theta_{m}}{\rho} J_1(\tilde{\rho})\boldsymbol{e}_y\\
\boldsymbol{E}_z&=&-iE_0\cdot \frac{\theta_{m}^2}{\rho} J_2(\tilde{\rho}) (\cos\varphi\boldsymbol{e_x}+\sin\varphi\boldsymbol{e_y})
\end{eqnarray}
and we defined the amplitude 
\begin{equation}
E_0=\frac{\mu\omega^2}{4\pi\epsilon_0^2c^2}\sqrt{\frac{n}{n'}}.
\end{equation}

\subsection{Estimation of the image position}
In the following we discuss the effect of the polarization of the dipole on the position of the image. For this, we discuss the case where the dipole has elliptical polarization in a plane that contains the imaging axis ($z$-axis). Without loss of generality, we assume the dipole to be polarized in a superposition of $\boldsymbol{\mu}=\boldsymbol{\mu}_x+ \epsilon \boldsymbol{\mu}_z$ where $\epsilon=\epsilon_r+i\epsilon_i$ is the complex valued ratio between longitudinal and transverse amplitude of the dipole. In particular we expect a strong displacement of the apparent position for the case where the dipole is circular polarized, \emph{i.e.}, $\epsilon_i=1$ and $\epsilon_r=0$. The intensity distribution in the image is given by the sum of the electric fields emitted by the two dipoles
\begin{eqnarray}
I_{dip}&=&|\boldsymbol{E}_x+ i \epsilon \boldsymbol{E}_z|^2\\\nonumber
&=&|\boldsymbol{E}_x|^2+\epsilon^2|\boldsymbol{E}_z|^z+(\epsilon^*\boldsymbol{E}_x\boldsymbol{E}_z^*+\epsilon\boldsymbol{E}_x^*\boldsymbol{E}_z).
\end{eqnarray}

Extended Figure 5 shows this distribution for different values of $\epsilon$ and NA. In order to obtain the position of the intensity distribution, we calculate the the centre of mass position $\boldsymbol{\rho}_{cms}$ of the image by
\begin{eqnarray}
\boldsymbol{\rho}_{cms}&=&\frac{\int dA \boldsymbol{\rho}I_{dip}}{\int dA I_{dip}}\\
&=& \frac{ \int_0^\infty \int_0^{2\pi} \rho\, d\rho\, d\varphi \left(
\begin{array}{c}
\rho \cos\varphi\\
\rho \sin\varphi
\end{array}
\right)I_{dip}}
{\int_0^\infty \int_0^{2\pi} \rho\, d\rho\, d\varphi I_{dip} }
\end{eqnarray}
Evaluating the above integrals can be done analytically and yields for the centre of mass position of the intensity distribution
\begin{equation}
\boldsymbol{\rho}_{cms}=\frac{2\epsilon_i}{k'(2+|\epsilon^2|\theta_{m}^2)}\frac{f'}{f}\left(\begin{array}{c} 1\\ 0\end{array}\right) \approx\frac{\epsilon_i}{k'}\frac{f'}{f}\left(\begin{array}{c} 1\\ 0\end{array}\right) \label{eqn_imagepos}
\end{equation}
where the last approximation is valid for $|\epsilon|\theta_{m}\ll 1$. In this limit and for $n'=1$ we obtain for the apparent position of the particle the ratio of the main axis of the elliptical polarization times $\lambda/2\pi$ times the magnification of our optical system $f'/f$. For circular polarization ($\epsilon_i=\pm1$, $\epsilon_r=0$) this yields the result expected from the wavefront argument in the previous chapter. Extended Figure 6 shows the displacement of the centre of mass for different values of $\epsilon$ and $NA$.

\subsection{Interference between dipoles}
If the point-like particle is emitting light that is a superposition of the emission of a dipole oriented along the optical axis (z) and a dipole orthogonal to the optical axis ($\perp$), then the total field is $\boldsymbol{E_{dip}}=\boldsymbol{E}_\perp+\boldsymbol{E}_z$ and the point-spread function of the emitter is given by 
\begin{eqnarray}
I_{dip}&=&|\boldsymbol {E_\perp}|^2+|\boldsymbol {E_z}|^2+2\Re(\boldsymbol{ E'_\perp}\cdot \boldsymbol{E_z^{*}}) \nonumber \\
&=& \left|E_\perp\frac{\pi d^2}{g}\right|^2 \cdot \frac{1}{\tilde\rho^2} \Big[ J_1^2(\tilde\rho)+|\epsilon|^2\frac{\theta_m^4}{\rho^2} J_2^2(\tilde\rho)+\nonumber\\ 
&& 2\Re(\epsilon)\frac{\theta_m^2}{\rho}\cos(\varphi-\varphi_\perp)J_1(\tilde\rho)J_2(\tilde\rho)\Big]\label{eq:intensity}
\end{eqnarray}
Here, $\varphi_\perp$ is the angle defining the polarization of the transverse polarized field and $\epsilon=i E_z/E_\perp$ is the (complex) ratio of the two field amplitudes of the original dipoles. As is obvious from Eq. \ref{eq:intensity}, interference only occurs when the two fields are at least partially in phase, \emph{i.e.}, when $\Re(\epsilon)\neq0$ , and is maximum when $\epsilon$ is fully real.

\section{Angular momentum conservation}

Electromagnetic radiation can carry angular momentum as spin angular momentum and orbital angular momentum. The first corresponds to the rotating electric and magnetic fields of circularly polarized radiation while the latter arises from the disposition of the wavefronts of the fields \citeSupplement{bliokh2015}. Spin and orbital angular momenta can transform into each other and be converted to mechanical angular momentum when interacting with matter \citeSupplement{beth1936,allen1992}. The apparent displacement on the position of the atom depending on the observed atomic transition can be understood as spin-orbit coupling of angular momentum of the single photons spontaneously emitted. To illustrate the effect, let us consider a photon emitted in a $\Delta m = +1$ or $\Delta m = -1$ electric dipole transition with wavelength $\lambda$, where $\Delta m = m_f - m_i$ is the difference in the magnetic quantum number of the final and initial electronic states of the atom, with the quantization axis oriented along $z$. In this case, the final angular momentum of the atom differs by a quantum $\hbar$ from the initial one, and for conservation of the total angular momentum this must be present in the emitted photon, as spin or orbital angular momentum. In general, for any direction, the expectation values of the spin and orbital angular momentum along $z$ of the photon are
\begin{align}
\langle s_z \rangle &= \Delta m \frac{2\cos ^2 \theta}{1+\cos ^2 \theta }\hbar, \\
\langle l_z \rangle &= \Delta m \frac{\sin ^2 \theta}{1+\cos ^2 \theta }\hbar 
\label{conservation}
\end{align}
respectively \citeSupplement{moeconservation}, where $\theta$ is the polar angle in spherical coordinates. For $\Delta m = \pm 1$ the expectation values of the spin and orbital angular momenta always add to $\mp\hbar$. A photon detected in the $z$ direction carries angular momentum solely in the form of spin angular momentum, \emph{i.e.}, it has right circular polarization $-\frac{1}{\sqrt{2}}(\hat{x}+i\hat{y})$ for a $\Delta m = +1$ transition and left circular polarization $\frac{1}{\sqrt{2}}(\hat{x}-i\hat{y})$ in the case of a $\Delta m = -1$ transition. For a photon detected in a direction $\hat{r}$ of the equatorial $xy$-plane ($\theta =\pi/2$) the polarization is linear, parallel to $\hat{r}\times \hat{z}$, irrespective of the $\Delta m = \pm 1$ transition, therefore the photon does not carry spin angular momentum. Conservation of the total angular momentum imposes that the photon carries an orbital angular momentum $l_z = \pm \hbar$.

Let now consider photons detected in the equatorial plane. In that plane, the wavefronts are spirals with opposite orientations. This angular variation of the phase is the signature of the presence of orbital angular momentum in the radiated field, as calculated in \citeSupplement{moeconservation}. The direction $\vec{k}$ of propagation of the photons, perpendicular to the wavefronts, is slightly tilted with respect to the radial direction, which cause the displacement of the intensity profile \citeSupplement{li2010macroscopic}. On the other hand, when detecting photons coming from the $\Delta m = \pm 1$ along the $\hat{z}$ axis, as the wavefront in this direction presents angular symmetry, they propagation of photons is not tilted. For detection in any other direction, the photon will carry a mixture of orbital angular momentum, as angular asymmetry in the wave front and spin angular momentum, as elliptical polarization.
For the detection of photons emitted by a $\Delta m =0$, photons do not carry any angular momentum, so they polarization is linear and the wavefronts are spherically symmetric for any direction, so displacement of the intensity profile does not occur. 
When considering also the imaging system, the angular momentum of the emitted field is not necessarily conserved, giving rise to additional apartment displacement in measured intensity profile in the image plane.

Conservation of angular momentum in single scatterers has being studied in the case of non-absorbing spherical particles, both in Mie and Rayleigh scattering \citeSupplement{schwartz2006}. It has been shown that the angular momentum is conserved separately in each direction \citeSupplement{schwartz2006}. If we consider an input field with right circular polarization, propagating along the $z$-axis towards the spherical particle, then the spin angular momentum angular distribution is given, in the Rayleigh case for the $z$ component by

\begin{align}
s_z(\theta)= \frac{3\epsilon_0 c}{16 \pi \omega } \frac{E^2_0}{r^2}\sigma_{sc}\cos^2(\theta),
\end{align}
where $\sigma_{sc}$ is the Rayleigh scattering cross-section. The expression for the orbital angular momentum is the complementary expression, $l_z = 1 - s_z$. These results are similar to the one found for the radiating dipole, meaning that for some directions, the angular momentum is present totally as spin, and for others as orbital angular momentum, and in general, a combination of both. The apparent displacement of the scatter can then be interpreted as the presence of orbital angular momentum.

\bibliographystyleSupplement{naturemag}
\bibliographySupplement{library}
\clearpage

\end{document}